\documentclass[longauth,structabstract]{aa}  
\usepackage{graphicx}
\usepackage{txfonts}
\begin{document}
   \title{Multi-site campaign for transit timing variations of WASP-12~b: possible detection of a long-period signal of planetary origin\thanks{Partly based on (1) data collected with the Nordic Optical Telescope, operated on the island of La Palma jointly by Denmark, Finland, Iceland, Norway, and Sweden, in the Spanish Observatorio del Roque de los Muchachos of the Instituto de Astrof\'{\i}sica de Canarias, (2) observations made at the Centro Astron\'omico Hispano Alem\'an (CAHA), operated jointly by the Max-Planck Institut f\"ur Astronomie and the Instituto de Astrof\'{\i}sica de Andaluc\'{\i}a (CSIC), (3) data collected with telescopes at the Rozhen National Astronomical Observatory, and (4) observations obtained with telescopes of the University Observatory Jena, which is operated by the Astrophysical Institute of the Friedrich-Schiller-University.}}

  \titlerunning{Multi-site campaign for transit timing variations of WASP-12~b}

   \author{G.~Maciejewski\inst{1},
		D.~Dimitrov\inst{2},
		M.~Seeliger\inst{3},
		St.~Raetz\inst{3},
		{\L}.~Bukowiecki\inst{1},
		M.~Kitze\inst{3},
		R.~Errmann\inst{3},
		G.~Nowak\inst{1},
		A.~Niedzielski\inst{1},
		V.~Popov\inst{2},
		C.~Marka\inst{3},
		K.~Go\'zdziewski\inst{1},
		R.~Neuh\"auser\inst{3},
		J.~Ohlert\inst{4,5},
		T.~C.~Hinse\inst{6},
		J.~W.~Lee\inst{6}, 
		C.-U.~Lee\inst{6},
		J.-N.~Yoon\inst{7},
		A.~Berndt\inst{3},
		H.~Gilbert\inst{3},
		Ch.~Ginski\inst{3},
		M.~M.~Hohle\inst{3},
		M.~Mugrauer\inst{3},
		T.~R\"oll\inst{3},
		T.~O.~B.~Schmidt\inst{3},
		N.~Tetzlaff\inst{3},
		L.~Mancini\inst{8},
		J.~Southworth\inst{9},
		M.~Dall'Ora\inst{10},
		S.~Ciceri\inst{8},
		R.~Zambelli\inst{11},
		G.~Corfini\inst{12},
 		H.~Takahashi\inst{13},
		K.~Tachihara\inst{14,15},
		J.~M.~Benk\H{o}\inst{16}, 
		K.~S\'arneczky\inst{16,17},
		Gy.~M.~Szabo\inst{16,17,18},
		T.~N.~Varga\inst{19},
		M.~Va\v{n}ko\inst{20},
		Y.~C.~Joshi\inst{21},
		W.~P.~Chen\inst{22}
          }

   \institute{Toru\'n Centre for Astronomy, Nicolaus Copernicus University, 
             Gagarina 11, PL--87100 Toru\'n, Poland\\
              \email{gm@astri.umk.pl}
	\and
		Institute of Astronomy, Bulgarian Academy of Sciences, 
		72 Tsarigradsko Chausse Blvd., 1784 Sofia, Bulgaria
        \and
              Astrophysikalisches Institut und Universit\"ats-Sternwarte, 
              Schillerg\"asschen 2--3, D--07745 Jena, Germany 
        \and
		Michael Adrian Observatorium, Astronomie Stiftung Trebur, Trebur, Germany
        \and
		University of Applied Sciences, Technische Hochschule Mittelhessen, Friedberg, Germany
        \and
		Advanced Astronomy and Space Science Division, 
		Korea Astronomy and Space Science Institute, Daejeon 305-348, Republic of Korea
       \and
 		Chungbuk National University Observatory, Cheongju 365-863, Republic of Korea
       \and
 		Max Planck Institute for Astronomy, K\"onigstuhl 17, 69117 Heidelberg, Germany
        \and
		Astrophysics Group, Keele University, Staffordshire, ST5 5BG, UK
        \and
		INAF - Astronomical Observatory of Capodimonte
		Salita Moiariello 16, 80131 - Napoli, Italy
	\and
		Societ\'a Astronomica Lunae, Castelnuovo Magra, 19030 - La Spezia, Italy
	\and
		Unione Astrofili Italiani, Sezione Stelle Variabili - GRAV
	\and
		Institute of Astronomy, The University of Tokyo,
		2-21-1 Osawa, Mitaka, Tokyo 181-0015, Japan
	\and
		Joint ALMA Observatory, Alonso de C{¥'o}rdova 3107, Vitacura, Santiago, Chile
	\and
		National Astronomical Observatory of Japan, 2-21-1, Osawa, Miaka, Tokyo, 181-8588, Japan
	\and
		MTA CSFK, Konkoly Observatory, Konkoly-Thege Miklo\'s \'ut 15-17, H-1121 Budapest, Hungary
	\and
		ELTE Gothard--Lend\"ulet Research Group, H-9700 Szombathely, Hungary
	\and
		Dept. of Exp. Physics \& Astronomical Observatory, University of Szeged, H-6720 Szeged, Hungary
	\and
		Department of Physics, E\"otv\"os University, P\'azm\'any P\'eter S\'et\'any 1/A, 
		1117 Budapest, Hungary
	\and
		Astronomical Institute, Slovak Academy of Sciences, 059 60 Tatransk\'a Lomnica, Slovakia		
	\and
		Aryabhatta Research Institute of Observational Sciences (ARIES),
		Manora Peak, Nainital 263129, India
        \and
		Institute of Astronomy, National Central University, 300 Jhongda Rd., Jhongli 32001, Taiwan
              }
   \authorrunning{G.~Maciejewski et al.}
   \date{Received ...; accepted ...}

 
  \abstract
   {}
   {The transiting planet WASP-12~b was identified as a potential target for transit timing studies because a departure from a linear ephemeris was reported in the literature. Such deviations could be caused by an additional planet in the system. We attempt to confirm the existence of claimed variations in transit timing and interpret its origin.}
   {We organised a multi-site campaign to observe transits by WASP-12~b in three observing seasons, using 0.5--2.6-metre telescopes.}
   {We obtained 61 transit light curves, many of them with sub-millimagnitude precision. The simultaneous analysis of the best-quality datasets allowed us to obtain refined system parameters, which agree with values reported in previous studies. The residuals versus a linear ephemeris reveal a possible periodic signal that may be approximated by a sinusoid with an amplitude of $0.00068\pm0.00013$ d and period of $500\pm20$ orbital periods of WASP-12~b. The joint analysis of timing data and published radial velocity measurements results in a two-planet model which better explains observations than single-planet scenarios. We hypothesize that WASP-12~b might be not the only planet in the system and there might be the additional 0.1~$M_{\rm{Jup}}$ body on a 3.6-d eccentric orbit. A dynamical analysis indicates that the proposed two-planet system is stable over long timescales.}
   {}

   \keywords{planetary systems -- stars: individual: WASP-12 -- planets and satellites: individual: WASP-12 b}

   \maketitle


\section{Introduction}\label{Introduction}

The transiting extrasolar planet WASP-12~b was found to be one of the most intensely irradiated planets (Hebb et al.~\cite{Hebb}). It orbits a G-type star on a tight orbit with a semimajor axis of $a_{\rm{b}}=0.0229\pm0.0008$ AU and an orbital period of only $\sim$1.09~d. The planet's proximity to the star results in a high equilibrium temperature of 2500~K, thus inducing numerous studies of the properties of the planetary atmosphere (Li et al.\ \cite{Li}; Lai et al.\ \cite{Lai}; Fossati et al.\ \cite{Fossatib}; Croll et al.\ \cite{Croll}; Crossfield et al.\ \cite{Crossfield}). In the discovery paper, the planet was found to have a mass of $m_{\rm{b}}=1.41\pm0.10$ $M_{\rm{Jup}}$ and an unexpectedly large radius of $R_{\rm{b}}=1.79\pm0.09$ $R_{\rm{Jup}}$. The host star was found to have an effective temperature of $T_{\rm{eff}}=6300^{+200}_{-100}$ K and a mass of $M_{*}=1.35\pm0.14$ $M_{\odot}$. Follow-up observations, primarily high-precision photometry, confirmed the large planetary radius (Maciejewski et al.\ \cite{Maciejewski11}, Chan et al.\ \cite{Chan}). The system parameters were also refined by Southworth (\cite{Southworth12}) who used high-precision literature data in a joint analysis. 

The orbital eccentricity of WASP-12~b, $e_{\rm{b}}=0.049\pm0.015$, was initially determined from the radial velocity (RV) measurements (Hebb et al.\ \cite{Hebb}). L\'opez-Morales et al.\ (\cite{Lopez}) observed occultations whose timing supports a non-circular orbit. On the other hand, further observations of occultations by Campo et al.\ (\cite{Campo11}) yielded mid-occultation times at orbital phases $0.5010\pm0.0006$ and $0.5006\pm0.0007$, consistent with a circular orbit. The same conclusion was reached by Husnoo et al.\ (\cite{Husnoo}) who analysed spectroscopic data enhanced by new RV measurements. Croll et al.\ (\cite{Croll}) observed occultations at a phase equal to $0.4998^{+0.0008}_{-0.0007}$. These authors re-analysed the available photometric and spectroscopic data and finally concluded that $e_{\rm{b}}$ is likely very close to zero. 

A possible non-circular orbit might drive the dissipation of tidal energy. This mechanism, in turn, might be responsible for heating up the planet's interior and bloating its radius in consequence (Miller et al.\ \cite{Miller}). The orbit of WASP-12~b is expected to be circularised on short timescales if the tidal dissipation constant is not significantly larger than a value typical for giant planets (Hebb et al.\ \cite{Hebb}; Ibgui et al.\ \cite{Ibgui}). Its eccentricity, if real, could be sustained by gravitational perturbations from an additional planet in the system (Hebb et al.\ \cite{Hebb}; Li et al.\ \cite{Li}). Such a perturbing body could affect the orbital motion of WASP-12~b, causing its transits to exhibit a departure from a linear ephemeris (Miralda--Escud\'e \cite{miralda02}; Schneider \cite{Schneider04}; Holman \& Murray \cite{holmanmurray05}; Agol et al.\ \cite{agoletal05}; Steffen et al.\ \cite{steffenetal07}). 

\begin{table*}
\caption{Observatories and instruments (with abbreviations identifying datasets) taking part in the campaign.} 
\label{table:1}      
\centering                  
\begin{tabular}{r l  l c c}      
\hline\hline                
\# & Observatory     & Telescope (abbreviation)/Instrument  & \diameter~(m) & $N_{\rm{tr}}$ \\
\hline 
1 & Observatorio del Roque de los Muchachos, La Palma (Spain) & Nordic Optical Telescope (NOT)/ALFOSC & 2.6 & 5 \\
2 & Calar Alto Astronomical Observatory (Spain) & 2.2-m reflector (Calar Alto)/CAFOS & 2.2 & 15 \\
3 & National Astronomical Observatory Rozhen (Bulgaria) & Ritchey-Chr\'etien-Coud\'e (2.0Rozhen)/CCD & 2.0 & 10 \\
 &  & Cassegrain (0.6Rozhen)/CCD & 0.6 & 5 \\
4 & Bohyunsan Optical Astronomy Observatory (South Korea) & Bohyunsan Optical Telescope (BOT)/CCD & 1.8 & 1 \\
5 & Osservatorio Astronomico di Bologna, Loiano (Italy) & Cassini Telescope (Loiano)/BFOSC & 1.5 & 2 \\
6 & Gunma Astronomical Observatory (Japan) & 1.5-m Ritchey-Chr\'etien (Gunma)/CCD & 1.5 & 2 \\
7 & Devasthal Observatory (India) & Devasthal Fast Optical Telescope (DFOT)/CCD & 1.3 & 1 \\
8 & Michael Adrian Observatory, Trebur (Germany) & T1T (Trebur)/CCD & 1.2 & 6 \\
9 & Mt.\ Lemmon Optical Astronomy Observatory (South Korea/USA) & 1.0-m reflector (Mt.\ Lemmon)/CCD & 1.0 & 3 \\
10 & Konkoly Observatory (Hungary) & Ritchey-Chr\'etien-Coud\'e (Konkoly)/CCD & 1.0 & 2 \\
11 & Lulin Observatory & Lulin One-metre Telescope (LOT)/CCD & 1.0 & 1 \\
12 & University Observatory Jena (Germany) & Schmidt Teleskop Kamera (Jena)/CCD & 0.9 & 5 \\
13 & Star\'a Lesn\'a Observatory (Slovakia) & 0.5-m reflector (St\'ara Lesn\'a)/CCD & 0.5 & 3 \\                       
\hline                                   
\end{tabular}
\tablefoot{\diameter\ is the diameter of a telescope's main mirror, and $N_{\rm{tr}}$ is the total number of complete or partial transit light curves.}
\end{table*}

The transit timing variation (TTV) method has been already used to confirm the planetary nature of systems with multiple transiting planets (e.g. Holman et al.\ \cite{HolmanKep9}) and to discover an additional planetary companion which cannot be detected with other techniques (Ballard et al.\ \cite{Ballard11}). All these discoveries are based on data acquired with the {\it Kepler} space telescope (Borucki et al.\ \cite{Borucki}) and, to our knowledge, no independently confirmed TTV signal has been detected from the ground so far\footnote{Here we mean TTV detection confirmed e.g.\ by RV or transit signals of the additional planet or by additional TTV observations of an independent group.}. Maciejewski et al.\ (\cite{Maciejewski11}) reported on a sign of a TTV signal for WASP-12~b and put upper constraints on a mass of a possible perturber. Being encouraged by this result, we organised a multi-site campaign for follow-up timing observations of transits of WASP-12~b.


\section{Observations and data reduction}\label{Observations}

During three consecutive observing seasons, spanning from the autumn of 2009 to the winter of 2012, we acquired 61 complete or partial transit light curves. We used fourteen 0.5--2.6-m telescopes located in thirteen observatories distributed around the world. A portion of data was obtained in collaboration with the Young Exoplanet Transit Initiative (YETI, Neuh\"auser et al.\ \cite{Neuhauser11}). A list of participating observatories and instruments (sorted by telescope mirror diameter) is presented in Table \ref{table:1}. To quantify the quality of each light curve, we used the photometric noise rate ($pnr$) defined by Fulton et al.\ (\cite{Fulton}) as
\begin{equation}
  pnr = \frac{rms}{\sqrt{\Gamma}}
\end{equation}
where the root mean square of the residuals, $rms$, is calculated from the light curve and a fitted model, and $\Gamma$ is the median number of exposures per minute. Observations were recorded in Coordinated Universal Time (UTC). To exclude systematic errors in the recorded times, local computer clocks were verified with the Network Time Protocol software, usually accurate to better than 0.1\,s. Observations were performed using the filters in which individual instruments were the most efficient (usually $R$). Most of telescopes were significantly defocused to reduce flat-fielding noise and decrease the amount of time lost due to CCD readout. Exposure times were kept fixed during each observing sequence. Table \ref{table:2} lists individual observations acquired with 2-m class telescopes while data obtained with smaller instruments are given in Table \ref{table:3}. Short descriptions of individual instruments, observation details, and data reduction are given in Sects.~\ref{NOT}--\ref{StaraLesna}. In further analysis, three amateur light curves (Sect.~\ref{Amateur}) were also included.

\onltab{2}{
\begin{table*}
\caption{Transit light curves observed with the $\geq$1.8-m telescopes.} 
\label{table:2}      
\begin{tabular}{r l r l c c c l l c}      
\hline\hline                
\# & Date UT     & Epoch        & Telescope & Filter & $\Gamma$ & $pnr$ & $T_{\rm{mid}}$ (BJD$_{\rm{TDB}}$)  & $O-C$ & HQ\\ 
 & & & & & & (mmag) &  $2450000+$ & (d)  & \\
\hline                        
1 & 2009 Nov 12 & 586 & 2.0Rozhen & $R_{\rm{Cousins}}$ & 4.60 & 0.78 & 5148.5514$^{+0.0010}_{-0.0010}$ & +0.0016 & --\\
2 & 2010 Nov 29 & 936 & 2.0Rozhen & $R_{\rm{Cousins}}$ & 1.77 & 1.40 & 5530.54705$^{+0.00070}_{-0.00077}$ & --0.00009 & --\\
3 & 2010 Dec 11 & 947 & 2.0Rozhen & $R_{\rm{Cousins}}$ & 1.77 & 0.79 & 5542.55271$^{+0.00028}_{-0.00029}$ & --0.00007 & yes\\
4 & 2011 Jan 03 & 968 & Calar Alto & $R_{\rm{Cousins}}$ & 0.83 & 0.70 & 5565.4718$^{+0.0011}_{-0.0011}$ & --0.0008 & --\\
5 & 2011 Jan 04 & 969 & Calar Alto & $R_{\rm{Cousins}}$ & 1.05 & 0.71 & 5566.56333$^{+0.00022}_{-0.00020}$ & --0.00070 & yes\\
6 & 2011 Feb 07 & 1000 & Calar Alto & $R_{\rm{Cousins}}$ & 0.98 & 0.76 & 5600.39791$^{+0.00026}_{-0.00026}$ & --0.00018 & yes\\
7 & 2011 Feb 08 & 1001 & Calar Alto & $R_{\rm{Cousins}}$ & 0.83 & 1.06 & 5601.48963$^{+0.00039}_{-0.00038}$ & +0.00013 & --\\
8 & 2011 Feb 19 & 1011 & Calar Alto & $R_{\rm{Cousins}}$ & 0.94 & 1.11 & 5612.4042$^{+0.0017}_{-0.0016}$ & +0.0005 & --\\
9 & 2011 Mar 02 & 1021 & Calar Alto & $R_{\rm{Cousins}}$ & 0.88 & 0.75 & 5623.3178$^{+0.0009}_{-0.0009}$ & --0.0001 & --\\
10 & 2011 Oct 07 & 1222 & Calar Alto & $R_{\rm{Cousins}}$ & 0.82 & 1.05 & 5842.69343$^{+0.00091}_{-0.00093}$ & --0.00010 & --\\
11 & 2011 Nov 10 & 1253 & Calar Alto & $R_{\rm{Cousins}}$ & 0.84 & 0.99 & 5876.52839$^{+0.00030}_{-0.00028}$ & +0.00081 & yes\\
12 & 2011 Nov 21 & 1263 & 2.0Rozhen & $R_{\rm{Cousins}}$ & 1.54 & 0.86 & 5887.44183$^{+0.00035}_{-0.00038}$ & +0.00004 & yes\\
13 & 2011 Nov 22 & 1264 & NOT & $R_{\rm{Bessel}}$ & 1.72 & 0.85 & 5888.53371$^{+0.00037}_{-0.00038}$ & +0.00050 & yes\\
14 & 2011 Nov 22 & 1264 & 2.0Rozhen & $R_{\rm{Cousins}}$ & 1.54 & 1.27 & 5888.53351$^{+0.00048}_{-0.00049}$ & +0.00030 & --\\
15 & 2011 Nov 23 & 1265 & NOT & $R_{\rm{Bessel}}$ & 1.82 & 0.57 & 5889.6253$^{+0.0009}_{-0.0009}$ & +0.0006 & --\\
16 & 2011 Nov 24 & 1266 & NOT & $R_{\rm{Bessel}}$ & 1.20 & 0.75 & 5890.71638$^{+0.00023}_{-0.00023}$ & +0.00033 & yes\\
17 & 2011 Dec 24 & 1293 & BOT & $R_{\rm{Cousins}}$ & 0.44 & 1.74 & 5920.18500$^{+0.00045}_{-0.00046}$ & +0.00059 & --\\
18 & 2011 Dec 27 & 1296 & 2.0Rozhen & $R_{\rm{Cousins}}$ & 1.23 & 0.62 & 5923.45846$^{+0.00020}_{-0.00020}$ & --0.00021 & yes\\
19 & 2012 Jan 18 & 1316 & 2.0Rozhen & $R_{\rm{Cousins}}$ & 1.23 & 0.82 & 5945.28657$^{+0.00079}_{-0.00078}$ & --0.00053 & yes\\
20 & 2012 Jan 19 & 1317 & Calar Alto & $R_{\rm{Cousins}}$ & 0.83 & 1.16 & 5946.37835$^{+0.00036}_{-0.00034}$ & --0.00016 & --\\
21 & 2012 Jan 19 & 1317 & 2.0Rozhen & $R_{\rm{Cousins}}$ & 1.23 & 1.00 & 5946.37872$^{+0.00034}_{-0.00033}$ & +0.00021 & --\\
22 & 2012 Jan 20 & 1318 & Calar Alto & $R_{\rm{Cousins}}$ & 0.82 & 0.87 & 5947.46973$^{+0.00021}_{-0.00022}$ & --0.00020 & yes\\
23 & 2012 Jan 21 & 1319 & Calar Alto & $R_{\rm{Cousins}}$ & 0.82 & 1.10 & 5948.56075$^{+0.00028}_{-0.00028}$ & --0.00060 & --\\
24 & 2012 Feb 01 & 1329 & NOT & $R_{\rm{Bessel}}$ & 1.54 & 0.51 & 5959.47548$^{+0.00018}_{-0.00018}$ & --0.00009 & yes\\
25 & 2012 Feb 02 & 1330 & NOT & $R_{\rm{Bessel}}$ & 1.58 & 0.45 & 5960.56687$^{+0.00033}_{-0.00040}$ & --0.00012 & yes\\
26 & 2012 Feb 12 & 1339 & Calar Alto & $R_{\rm{Cousins}}$ & 0.83 & 0.72 & 5970.38980$^{+0.00029}_{-0.00028}$ & +0.00003 & yes\\
27 & 2012 Feb 13 & 1340 & Calar Alto & $R_{\rm{Cousins}}$ & 0.83 & 0.89 & 5971.48128$^{+0.00029}_{-0.00029}$ & +0.00009 & yes\\
28 & 2012 Feb 23 & 1349 & 2.0Rozhen & $R_{\rm{Cousins}}$ & 1.23 & 0.85 & 5981.30398$^{+0.00094}_{-0.00093}$ & --0.00000 & yes\\
29 & 2012 Feb 24 & 1350 & Calar Alto & $R_{\rm{Cousins}}$ & 0.88 & 0.87 & 5982.39522$^{+0.00047}_{-0.00046}$ & --0.00018 & yes\\
30 & 2012 Feb 25 & 1351 & Calar Alto & $R_{\rm{Cousins}}$ & 0.87 & 0.94 & 5983.48682$^{+0.00029}_{-0.00028}$ & +0.00000 & yes\\
31 & 2012 Mar 18 & 1371 & 2.0Rozhen & $R_{\rm{Cousins}}$ & 1.54 & 0.87 & 6005.31541$^{+0.00038}_{-0.00037}$ & +0.00017 & yes\\
\hline                                   
\end{tabular}
\tablefoot{$\Gamma$ is the median number of exposures per minute, $pnr$ is the photometric noise rate, $T_{\rm{mid}}$ is the mid-transit time, $O-C$ is the difference from the linear ephemeris given in Sect.~\ref{Timing}, and HQ flags light curves used in the joint analysis (Sect.~\ref{System}).}
\end{table*}
} 

\onltab{3}{
\begin{table*}
\caption{Transit light curves observed with the $<$1.8-m telescopes.} \label{table:3}      
\begin{tabular}{r l r l c c c l l}      
\hline\hline                
\# & Date UT     & Epoch        & Telescope & Filter & $\Gamma$ & pnr & $T_{\rm{mid}}$ (BJD$_{\rm{TDB}}$)  & $O-C$ \\ 
 & & & & & & (mmag) & $2450000+$ & (d) \\
\hline                        
1 & 2009 Nov 11 & 585 & Loiano & Gunn-$r$ & 0.49 & 1.04 & 5147.45861$^{+0.00042}_{-0.00042}$ & +0.00020 \\
2 & 2009 Nov 22 & 595 & 0.6Rozhen & $R_{\rm{Bessel}}$ & 0.49 & 2.91 & 5158.3734$^{+0.0013}_{-0.0012}$ & +0.0008 \\
3 & 2009 Nov 23 & 596 & 0.6Rozhen & $R_{\rm{Bessel}}$ & 0.49 & 2.89 & 5159.46377$^{+0.00095}_{-0.00088}$ & --0.00026 \\
4 & 2009 Nov 24 & 597 & 0.6Rozhen & $R_{\rm{Bessel}}$ & 0.49 & 3.10 & 5160.55549$^{+0.00070}_{-0.00067}$ & +0.00003 \\
5 & 2009 Dec 26 & 626 & Gunma & $R_{\rm{Cousins}}$ & 1.00 & 1.93 & 5192.20540$^{+0.00049}_{-0.00050}$ & --0.00127 \\
6 & 2010 Jan 18 & 647 & Gunma & $R_{\rm{Cousins}}$ & 1.00 & 1.81 & 5215.12869$^{+0.00080}_{-0.00082}$ & +0.00218 \\
7 & 2010 Feb 10 & 668 & LOT & $R_{\rm{Bessel}}$ & 1.15 & 2.22 & 5238.04586$^{+0.00062}_{-0.00066}$ & --0.00048 \\
8 & 2010 Mar 09 & 693 & St\'ara Lesn\'a & $I_{\rm{Cousins}}$ & 0.97 & 2.58 & 5265.3313$^{+0.0015}_{-0.0016}$ & --0.0006 \\
9 & 2010 Oct 24 & 903 & Trebur & $R_{\rm{Bessel}}$ & 1.23 & 1.70 & 5494.53018$^{+0.00048}_{-0.00047}$ & --0.00007 \\
10 & 2010 Oct 24 & 903 & 0.6Rozhen & $R_{\rm{Bessel}}$ & 0.65 & 2.37 & 5494.53061$^{+0.00063}_{-0.00062}$ & +0.00035 \\
11 & 2011 Jan 13 & 977 & 0.6Rozhen & $R_{\rm{Bessel}}$ & 0.65 & 2.99 & 5575.29703$^{+0.00079}_{-0.00079}$ & +0.00163 \\
12 & 2011 Jan 27 & 990 & Konkoly & $R_{\rm{Cousins}}$ & 0.95 & 1.63 & 5589.48296$^{+0.00092}_{-0.00094}$ & --0.00091 \\
13 & 2011 Jan 28 & 991 & Trebur & $R_{\rm{Bessel}}$ & 0.87 & 2.52 & 5590.57597$^{+0.00100}_{-0.00092}$ & +0.00067 \\
14 & 2011 Feb 05 & 998 & DFOT & $R_{\rm{Cousins}}$ & 0.57 & 1.14 & 5598.21556$^{+0.00036}_{-0.00036}$ & +0.00032 \\
15 & 2011 Feb 07 & 1000 & Trebur & $R_{\rm{Bessel}}$ & 0.87 & 1.90 & 5600.39720$^{+0.00075}_{-0.00083}$ & --0.00088 \\
16 & 2011 Feb 07 & 1000 & St\'ara Lesn\'a & $R_{\rm{Cousins}}$ & 1.51 & 2.28 & 5600.3978$^{+0.0010}_{-0.0010}$ & --0.0003 \\
17 & 2011 Feb 08 & 1001 & Trebur & $R_{\rm{Bessel}}$ & 0.88 & 2.01 & 5601.48999$^{+0.00086}_{-0.00088}$ & +0.00048 \\
18 & 2011 Feb 08 & 1001 & Jena & $R_{\rm{Bessel}}$ & 1.04 & 2.03 & 5601.49023$^{+0.00076}_{-0.00078}$ & +0.00073 \\
19 & 2011 Mar 02 & 1021 & Jena & $R_{\rm{Bessel}}$ & 1.25 & 2.15 & 5623.31864$^{+0.00052}_{-0.00054}$ & +0.00072 \\
20 & 2011 Mar 03 & 1022 & Jena & $R_{\rm{Bessel}}$ & 0.82 & 2.77 & 5624.41007$^{+0.00086}_{-0.00082}$ & +0.00073 \\
21 & 2011 Mar 03 & 1022 & St\'ara Lesn\'a & $R_{\rm{Cousins}}$ & 1.48 & 2.86 & 5624.4104$^{+0.0011}_{-0.0014}$ & +0.0010 \\
22 & 2011 Oct 07 & 1222 & Loiano & Gunn-$r$ & 0.45 & 1.79 & 5842.6942$^{+0.0013}_{-0.0013}$ & +0.0007 \\
23 & 2011 Nov 21 & 1263 & Trebur & $R_{\rm{Bessel}}$ & 0.88 & 1.95 & 5887.44184$^{+0.00062}_{-0.00061}$ & +0.00006 \\
24 & 2011 Nov 21 & 1263 & Jena & $R_{\rm{Bessel}}$ & 2.04 & 1.83 & 5887.44237$^{+0.00074}_{-0.00075}$ & +0.00059 \\
25 & 2011 Nov 22 & 1264 & Jena & $R_{\rm{Bessel}}$ & 2.57 & 1.90 & 5888.53325$^{+0.00057}_{-0.00059}$ & +0.00005 \\
26 & 2011 Dec 05 & 1276 & Konkoly & $R_{\rm{Cousins}}$ & 0.33 & 2.58 & 5901.63002$^{+0.00078}_{-0.00083}$ & --0.00023 \\
27 & 2011 Dec 08 & 1278 & Mt.\ Lemmon & $R_{\rm{Cousins}}$ & 0.82 & 2.42 & 5903.81351$^{+0.00093}_{-0.00093}$ & +0.00041 \\
28 & 2011 Dec 31 & 1299 & Mt.\ Lemmon & $R_{\rm{Cousins}}$ & 0.82 & 1.67 & 5926.73339$^{+0.00092}_{-0.00089}$ & +0.00046 \\
29 & 2012 Feb 14 & 1311 & Mt.\ Lemmon & $R_{\rm{Cousins}}$ & 0.82 & 1.70 & 5939.8286$^{+0.0018}_{-0.0019}$ & --0.0014 \\
30 & 2012 Mar 19 & 1372 & Trebur & $R_{\rm{Bessel}}$ & 0.88 & 2.03 & 6006.40642$^{+0.00040}_{-0.00040}$ & --0.00025 \\
\hline                                   
\end{tabular}
\tablefoot{$\Gamma$ is the median number of exposures per minute, $pnr$ is the photometric noise rate, $T_{\rm{mid}}$ is the mid-transit time, and $O-C$ is the difference from the linear ephemeris given in Sect.~\ref{Timing}.}
\end{table*}
} 

\begin{figure*}
  \centering
  \includegraphics[width=18cm]{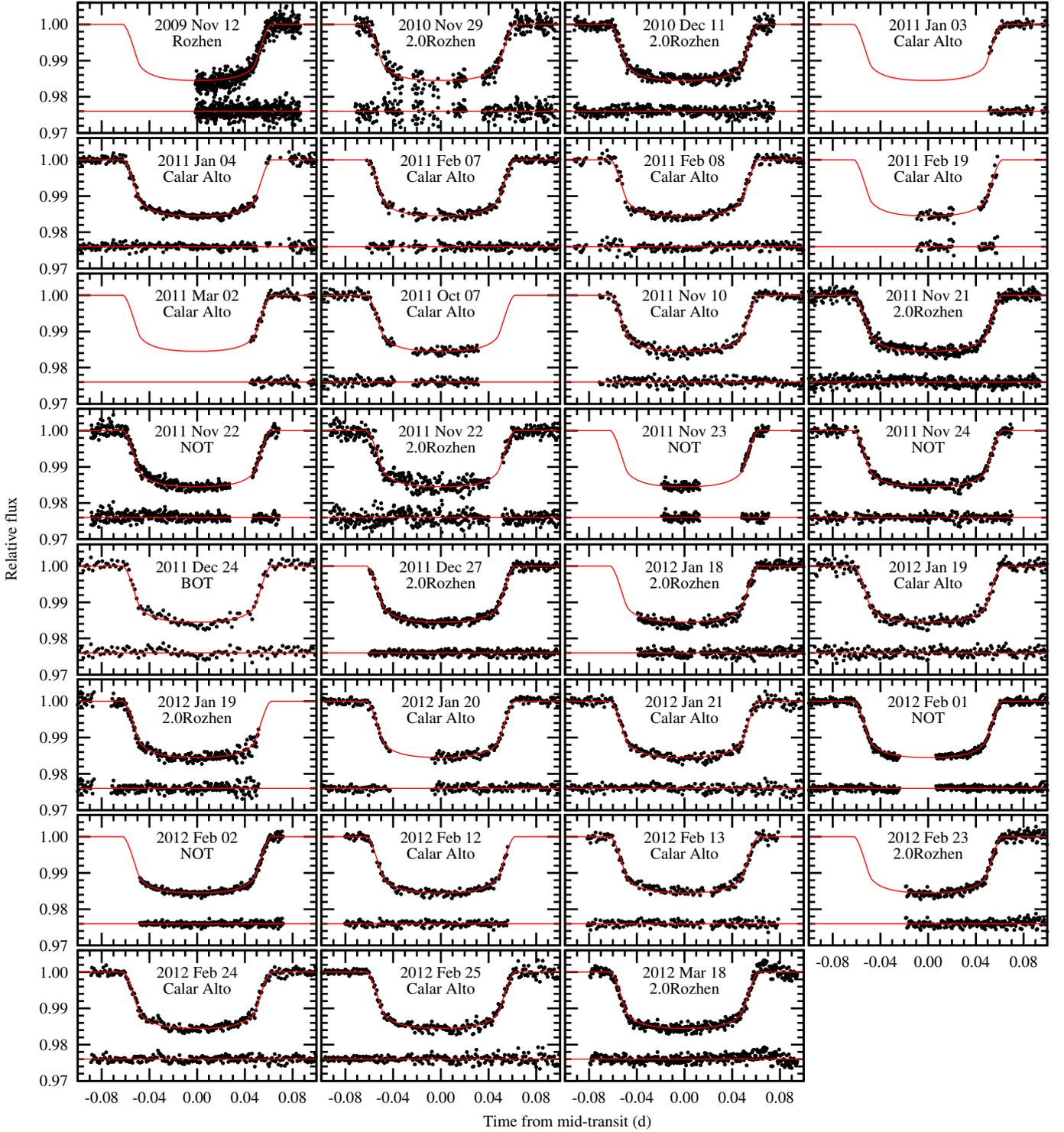}
  \caption{Light curves for transits of WASP-12~b observed with 2-m class telescopes. The residuals are shown in bottom plots.}
  \label{fig:hqlc}
\end{figure*}

\begin{figure*}
  \centering
  \includegraphics[width=18cm]{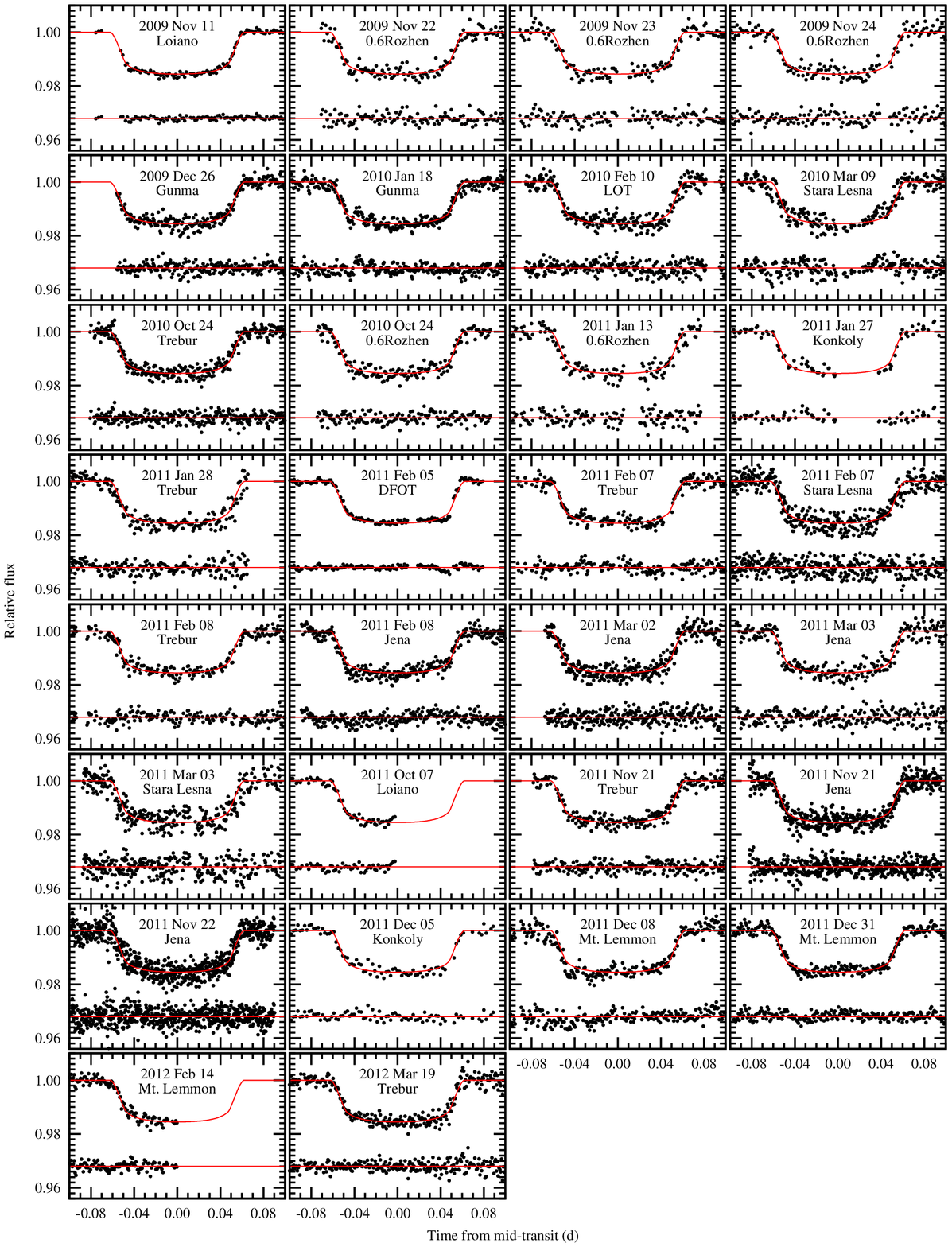}
  \caption{The same as Fig.~\ref{fig:hqlc} but for 1.5-m and smaller telescopes.}
  \label{fig:smalllc}
\end{figure*}

\subsection{Nordic Optical Telescope}\label{NOT}

Five light curves were acquired in the 2011/12 observing season with the Nordic Optical Telescope (NOT) at the Observatorio del Roque de los Muchachos, La Palma (Spain), through the P44-102 observing programme (OPTICON 2011B/003). The Andalucia Faint Object Spectrograph and Camera (ALFOSC) was used in imaging mode, giving a field of view (FoV) of $6\farcm4 \times 6\farcm4$. The CCD was binned $2\times2$ to obtain faster readout. During the first run (2011 Nov 22-24) we acquired two complete light curves and one partial light curve, for which some of the data were lost due to bad weather. The second run (2012 Feb 1-2) allowed us to observe two almost complete transits. The beginning of the event on Feb 2 was lost due to bad weather. Twenty-minute gaps in light curves on 2011 Nov 22 and 2012 Feb 1 are due to the target passing through zenith, causing rapid motion of the field derotator.  Autoguiding guaranteed that stellar centroids did not change their locations on the detector matrix during each night. 

The photometric measurements were performed with a custom-made pipeline employing differential aperture photometry and standard CCD reduction procedures (de-biasing and flat-fielding with sky flats). Only the nearest comparison star was used to minimise possible flat-field systematics induced by the field rotator (e.g.\ Moehler et al.\ \cite{Moehler}). A number of apertures were used and the light curves with the smallest photometric scatter were taken as the final ones.

\subsection{Calar Alto 2.2-m telescope}\label{CAHA2}

We observed 15 transits with the 2.2-m telescope at the Calar Alto Observatory (Spain). Most of these observations cover complete events and only four are partial due to bad weather. In the 2010/11 observing season, ten nights were awarded within the programme F11-2.2-008. We gathered usable data in six nights; the other four nights were completely lost due to bad weather. In the 2011/12 season, we got seven nights (programme F12-2.2-009) and in all of them acquired complete transit light curves. Two additional light curves were obtained on 2011 Oct 7 and Nov 10 as backup observations for the programme H11-2.2-011. 

The Calar Alto Faint Object Spectrograph (CAFOS) was used in imaging mode. The original FoV was windowed to $5\farcm6 \times 5\farcm8$ and $2\times2$ binning was applied to shorten the read-out time. The telescope was autoguided during all nights except 2012 Feb 24, when the observers were unable to use the autoguider due to significant telescope defocusing. The data reduction procedure was similar to that in Sect.~\ref{NOT}.

\subsection{2.0-m and 0.6-m telescopes at the National Observatory Rozhen}\label{ROZ}

The 2.0-m Ritchey-Chr\'etien-Coud\'e (RCC) telescope at the Bulgarian National Astronomical Observatory Rozhen was used to observe six complete and four partial transits. During the 2009/10 and 2010/11 seasons, a VersArray 1300B CCD camera ($1340 \times 1300$ pixels, 20 $\mu$m pixel size, resolution $0\farcs258$ per pixel) with a FoV of $5\farcm8 \times 5\farcm6$ was used as a detector. The missing part in the beginning of the event on 2009 Nov 12 as well as gaps on 2010 Nov 29 were caused by bad weather.

During the 2011/12 season, the same CCD camera was attached to the telescope through the two-channel focal reducer FoReRo-2, giving a resolution of 0\farcs74 per pixel and a FoV of about $15\arcmin$ in diameter. Gaps in data on 2011 Nov 22 and 2012 Jan 19 were caused by bad weather conditions. The beginnings of the events on 2011 Dec 27, 2012 Jan 18, and 2012 Feb 23 were lost due to technical problems with the telescope. Red noise is noticeable in some of the light curves. This effect is principally caused by observing without autoguiding, as the telescope's autoguider is unusable in the focal reducer mode.  

In addition, we acquired five transit light curves with the 0.6-m Cassegrain photometric telescope equipped with a FLI PL09000 CCD camera ($3056 \times 3056$ pixels, 12 $\mu$m pixel size, resolution $0\farcs334$ per pixel, FoV: $17\arcmin \times 17\arcmin$).  

Standard \textsc{IDL} procedures (adapted from \textsc{DAOPHOT}) were used for data reduction (de-biasing and flat-fielding with sky flats) and performing differential aperture photometry. Using the method of Everett \& Howell (\cite{everetthowell01}), several stars (4 to 6) with photometric precision better than 4\,mmag (6\,mmag for the 0.6-m telescope) were selected to create an artificial comparison star used for the differential photometry.

\subsection{Bohyunsan Optical Telescope}\label{BOT}

The transit light curve on 2012 Dec 24 was obtained with the Bohyunsan Optical Telescope (BOT) at the Bohyunsan Optical Astronomy Observatory (BOAO). The telescope is an alt-az 1.8-m (f/80) reflector operated by the Korea Astronomy and Space Science Institute (KASI), equipped with the BOAO 4k CCD ($15 \times 15$ $\mu$m) imaging instrument. An available FoV is $14\farcm6 \times 14\farcm6$ with a scale of $0\farcs21$ per pixel. We used the $2 \times 2$ binning mode and windowed the FoV to increase the sampling rate. The autoguider system was used and pointing errors were found to be of the order of a few tenths of a pixel per exposure. Observations were reduced in the same way as described in Sect.~\ref{NOT}.

\subsection{Cassini Telescope}\label{CassTel}

A complete transit of WASP-12 b was observed on 2009 Nov 11 and a partial one on 2011 Oct 7 with the 1.52-m Cassini Telescope at the Astronomical Observatory of Bologna in Loiano (Italy). This telescope was already successfully utilised to follow up several transiting planets (Southworth et al.\ \cite{southworth2010}, \cite{southworth2012a}, \cite{southworth2012b}; Mancini et al.\ \cite{mancini2012}). It is equipped with the Bologna Faint Object Spectrograph \& Camera (BFOSC), an instrument built to allow, with a simple configuration change, the acquisition of both images and spectra. The detector is a back-illuminated EEV LN/1300-EB/1 CCD with $1300 \times 1340$ pixels, giving a FoV of $13\farcm0 \times 12\farcm6$ at a scale of $0\farcs58$ per pixel.

The observations were analysed following standard methods. We created master bias and flat-field images by median-combining sets of bias images and sky flats, and used them to correct the science images. Aperture photometry was performed using the \textsc{IDL/ASTROLIB} implementation of \textsc{DAOPHOT} (Southworth et al.\ \cite{southworth2009a}). The apertures on the target star were placed by eye and their sizes were set to a wide range of values to find those which gave photometry with the lowest scatter compared to a fitted model. Differential photometry was finally performed using an optimal ensemble of comparison stars.

\subsection{1.5-m Gunma Astronomical Observatory telescope}\label{Gunma}

Transits on 2009 Dec 26 and 2010 Jan 18 were observed with the 1.5-m Ritchey-Chr\'etien telescope at the Gunma Astronomical Observatory (Japan). An Andor DW432 CCD camera ($1250 \times 1152$ pixels) was used as a detector providing a $12\farcm5 \times 11\farcm5$ FoV.  

The photometric data were reduced following standard procedures including subtraction of a median dark frame and dividing by a sky flat field. The relative magnitudes were derived with the aperture photometry method implemented in the customised software pipeline developed for the Semi-Automatic Variability Search sky survey (Niedzielski et al.\ \cite{niedzielskietal03}). An artificial comparison star was generated iteratively using 20--30\% of the unsaturated stars with the lowest scatter. A number of aperture radii were tested and the aperture which produced light curves with the smallest scatter was finally chosen.

\subsection{Devasthal Fast Optical Telescope}\label{DFOT}

The transit on 2011 Feb 5 was observed with the 1.3-m Devasthal Fast Optical Telescope (DFOT, Sagar et al.\ \cite{Sagar12}) located at the Devasthal Observatory (India). A set of 2770 exposures was recorded with an exposure time of 5\,s. The data reduction was done as described in detail in Joshi et al.\ (\cite{Joshi09}). To improve the signal-to-noise ratio, subsets of 20 single frames were co-added to produce the final light curve.

\subsection{1.2-m Michael Adrian Observatory telescope}\label{Trebur}

The 1.2-m T1T telescope at the Michael Adrian Observatory in Trebur (Germany) was used to observe four and two complete transits in the 2010/11 and 2011/12 seasons, respectively. The telescope has a Cassegrain optical system and is equipped with a $3072\times2048$-pixel SBIG STL-6303 CCD camera (FoV: $10\farcm0 \times 6\farcm7$). The light curve acquired on 2011 Jan 28 was affected by high airmass at the end of the run. Observations on 2011 Feb 7 were interrupted by occasional passing clouds. Observations were reduced in the same way as described in Sect.~\ref{Gunma}.

\subsection{Mount Lemmon Optical Telescope}\label{MLOT}

Two complete (2011 Dec 8 and 2012 Jan 1) and one partial (2012 Jan 13) transit light curves were obtained using the 1.0-m reflector operated by KASI at the Mount Lemmon Optical Astronomy Observatory (LOAO) in Arizona, USA. For all observations we used the ARC 4k CCD instrument in the $2\times2$ binning mode. The CCD has a pixel size of $15 \times 15$ $\mu$m, a scale of 0\farcs41 per pixel, and a FoV of $28\arcmin \times 28\arcmin$. During the transits the weather was clear but not photometric with partial clouds early in the evenings and later clearing up during the runs. Observations were reduced in the same way as described in Sect.~\ref{Gunma}.

\subsection{Konkoly Observatory}\label{Konkoly}

Transits on 2011 Jan 27 and 2011 Dec 5 were observed with the 1.0-m RCC Telescope at the Piszk\'estet\H{o} Station of the Konkoly Observatory (Hungary), equipped with a $1340\times1300$-pixel PI VersArray 1300b NTE CCD (FoV: $7\arcmin\times7\arcmin$, image scale $0\farcs32$ per pixel). Flat-fielding using sky flats, dark correction, and aperture photometry was performed with standard \textsc{IRAF} routines and a dedicated \textsc{gnu-r} script as described in Szab\'o et al.\ (\cite{SzaboH13}).

\subsection{Lulin One-meter Telescope}\label{LOT}

A transit on 2010 Feb 10 was observed with the Lulin One-meter Telescope (LOT) at the Lulin Observatory, operated by the National Central University of Taiwan. A VersArray:1300B CCD camera with a FoV of $11\arcmin\times11\arcmin$ was used as a detector. Observations were reduced in the same way as described in Sect.~\ref{Gunma}.

\subsection{University Observatory Jena}\label{Jena}

Five light curves were acquired with the 0.9/0.6-m Schmidt telescope at the University Observatory Jena in Gro{\ss}schwabhausen near Jena (Germany). The CCD-imager Schmidt Teleskop Kamera (STK, Mugrauer \& Berthold \cite{mugrauer}) has a $52\farcm8 \times 52\farcm8$ FoV which allows numerous comparison stars to be observed simultaneously. 

Data acquired on 2011 Feb 8, 2011 Mar 2 and 2011 Mar 3 were reduced as described in Sect.~\ref{Gunma}. The remaining the nights (2011 Nov 21 and 22) were reduced with the \textsc{IRAF} routines after subtracting a dark frames and applying flat-field correction with sky flats. After applying aperture photometry, differential magnitudes were calculated using the algorithm described in Broeg et al.\ (\cite{Broeg}). This algorithm calculates an artificial standard star out of all stars in the field, weighted by their magnitude errors, and compares the brightness of each star, including our target star WASP-12, against this standard star.

\subsection{Star\'a Lesn\'a Observatory}\label{StaraLesna}

Three light curves were obtained at the Star\'a Lesn\'a observatory in Slovakia. Observations were performed with the 0.5-m Newtonian telescope, on which an SBIG ST10 MXE CCD camera with 2184$\times$1472 6.8-$\mu$m pixels was mounted. The scale was 0\farcs56 per pixel, corresponding to a FoV of $24\arcmin \times 16\arcmin$.

The light curve obtained on 2010 Mar 9 exhibits correlated noise because observations were acquired without autoguiding. Standard correction procedures (bias, dark, and flat-fielding) and then aperture photometry was performed with the \textsc{C-munipack} package\footnote{http://c-munipack.sourceforge.net}. To generate an artificial comparison star, at least 20--30\% of stars with the lowest photometric scatter were selected iteratively from the field stars brighter than 2.5--3 mag below saturation level. To measure instrumental magnitudes, various aperture radii were used. The aperture which gave the smallest magnitude scatter was applied to generate a final light curve.

\subsection{Amateur light curves}\label{Amateur}

\begin{figure}
  \centering
  \includegraphics[width=9cm]{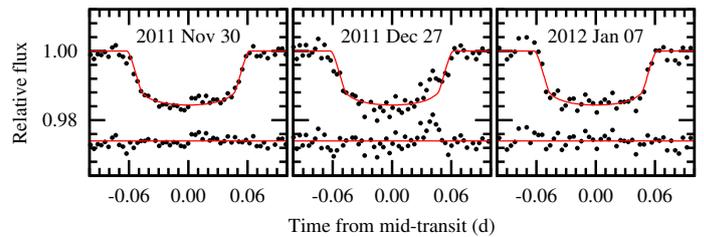}
  \caption{Additional amateur light curves obtained with 20-cm reflecting telescopes. The mid-transit times were found to be BJD$_{\rm{TDB}}$ $2455865.6130\pm0.0012$ (2011 Oct 30), $2455923.4609\pm0.0026$ (2011 Dec 27), and $2455934.3725\pm0.0009$ (2012 Jan 07).}
  \label{fig:amalc}
\end{figure}

Light curves for transits on 2011 Oct 30 and 2012 Jan 7 were acquired at Castelnuovo Magra (Italy) with a Meade LX200 GPS 10\,inch telescope equipped with an f/6.3 focal reducer and an SBIG ST8 XME CCD camera. Science frames were taken through the Baader Yellow 495 Longpass filter with an exposure time of 300\,s. The CCD frames were acquired with the \textsc{MaxIm DL 4} software and then reduced with \textsc{MaxIm DL 5}. The data reduction procedure included removing dark frames, flat-fielding, and aperture photometry against nearby bright stars.

Observations of the transit on 2011 Dec 27 were acquired with a CCD camera based on a Sony ICX429ALL sensor, mounted on an 0.2-m Newtonian telescope located near Lucca (Italy). The data were also collected through the Baader Yellow 495 passband filter and an exposure time of 260\,s was used. The equatorial fork mount of the telescope was guided by a modified Webcam coupled with a 60 mm telephoto lens. Images were acquired by the \textsc{AstroArt v4} software and a standard data reduction procedure including aperture photometry was performed with the \textsc{MaxIm DL 5} programme. To generate an artificial comparison star, eight reference stars of magnitudes close to that of WASP-12 were used.

\subsection{Light curve pre-processing}\label{Preprocessing}

Even the best-quality light curves may be affected by photometric trends caused, for instance, by colour difference between the target and comparison stars and changes of airmass during a run. If they are not removed properly they may seriously affect transit timing. To remove trends, a model that consists of a transit signal adopted from Maciejewski et al.\ (\cite{Maciejewski11}) and a second-order polynomial was fitted to each light curve analysed in this work. We used the {\sc jktebop} code (Southworth et al. \cite{Southwortha, Southworthb}), which is based on the {\sc ebop} program (Etzel \cite{Etzel}; Popper \& Etzel \cite{Popper}) and allows photometric trends to be approximated with a polynomial up to order five. A best-fitting trend was subtracted from each light curve and then magnitudes were transformed into fluxes normalised to unit out-of-transit flux level.


\section{Results}\label{Results}

\subsection{System parameters}\label{System}

The Transit Analysis Package\footnote{http://ifa.hawaii.edu/users/zgazak/IfA/TAP.html} ({\sc tap}, version 2.104,  Gazak et al.\ \cite{Gazak11}) was used to analyse transit light curves and refine system parameters. TAP uses the MCMC (Markov Chain Monte Carlo) approach, with the Metropolis-Hastings algorithm and a Gibbs sampler, to determine model parameters. Wavelet-based techniques are employed to account for correlated noise while estimating error values. It has been shown that this approach provides the most reliable parameter results and error estimates (e.g.\ Hoyer et al.\ \cite{Hoyer12}). The code employs the quadratic limb-darkening (LD) law to model the distribution of the flux on a stellar disc.

\begin{figure}
  \centering
  \includegraphics[width=9cm]{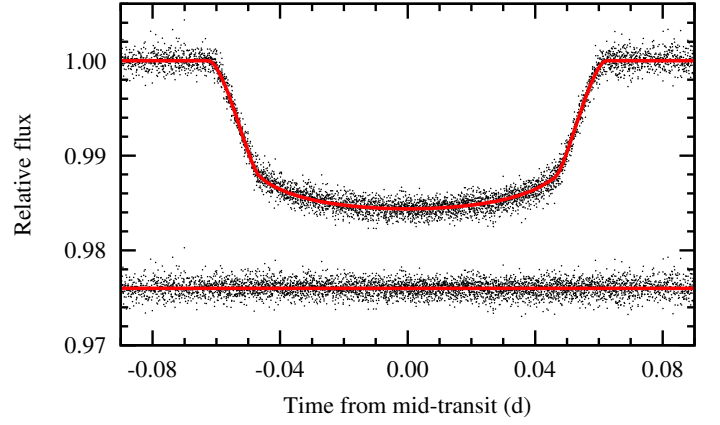}
  \caption{Best-fitting model of WASP-12~b's transit (red line) laid over the 19 best-quality, phase-folded light curves.}
  \label{fig:best}
\end{figure}

To refine the system parameters, the best-quality light curves acquired with 2-m class telescopes were selected for a simultaneous fit. The selection criteria included sub-millimagnitude precision of a dataset ($pnr < 1.0$ mmag) and transit phase completeness higher than $75\%$, including at least an ingress or/and egress preceded or followed by out-of-transit observations and a flat-bottom phase. Nineteen light curves were found to fulfil these criteria, including observations on 2010 Feb 02 from Maciejewski et al.\ (\cite{Maciejewski11}). We ran ten MCMC chains, each containing $10^6$ steps. The individual chains were combined to get final posteriori probability distributions. The following parameters were linked together for all light curves: orbital period $P_{\rm{b}}$, orbital inclination $i_{\rm{b}}$, semimajor-axis scaled by stellar radius $a_{\rm{b}}/R_{*}$, planetary to stellar radii ratio $R_{\rm{b}}/R_{*}$, and linear ($u$) and quadratic ($v$) LD coefficients. The mid-transit times, airmass slopes, and flux offsets were allowed to vary separately for individual light curves. A circular orbit was assumed. An analysis of relations between the parameters reveals significant correlation or anti-correlations (with the Pearson correlation coefficients ranging from 0.71 to 0.96) between $i_{\rm{b}}$ and $a_{\rm{b}}/R_{*}$, $i_{\rm{b}}$ and $R_{\rm{b}}/R_{*}$, $a_{\rm{b}}/R_{*}$ and $R_{\rm{b}}/R_{*}$, $R_{\rm{b}}/R_{*}$ and $v$, and $u$ and $v$.
The best-fitting parameters were determined by taking the median value of marginalised posteriori probability distributions which were found to be unimodal. The 15.9, and 85.1 percentile values of these distributions were taken as upper and lower 1\,$\sigma$ errors. The results are presented in Table~\ref{table:pars} where the impact parameter (defined as $b=\frac{a_{\rm{b}}}{R_{*}}\cos{i_{\rm{b}}}$) is also given. The final model of WASP-12~b's transit signature is plotted in Fig.~\ref{fig:best}.  

\begin{table}
\caption{Parameters of the WASP-12 system from the joint analysis of 19 sub-millimagnitude-precision light curves and the refined linear ephemeris.} 
\label{table:pars}                       
\centering                  
\begin{tabular}{l c}      
\hline\hline                
Parameter & Value\\ 
\hline 
Orbital inclination, $i_{\rm{b}}$                                 & $82\fdg96^{+0\fdg50}_{-0\fdg44}$  \\
Scalled semimajor-axis, $a_{\rm{b}}/R_{*}$              & $3.033^{+0.022}_{-0.021}$  \\
Planetary to stellar radii ratio, $R_{\rm{b}}/R_{*}$              & $0.1173^{+0.0005}_{-0.0005}$  \\
Transit parameter, $b$                                            & $0.355^{+0.028}_{-0.025}$  \\
Linear LD coefficient, $u$                                        & $0.301^{+0.041}_{-0.040}$  \\
Quadratic LD coefficient, $v$                                     & $0.24^{+0.08}_{-0.08}$  \\
Observed planetary radius $R_{\rm{b}}$ ($R_{\rm{Jup}}$) & $1.86 \pm 0.09$ \\
Effective planetary radius, $R_{\rm{b}}^{\rm{eff}}$ ($R_{\rm{Jup}}$) & $1.90 \pm 0.09$ \\
Stellar density, $\rho_{*}$  ($\rho_{\odot}$) & $0.315\pm0.007$ \\
Orbital period, $P_{\rm{b}}$ (d)         &          $1.0914209 \pm 0.0000002$\\
Cycle-zero transit time, $T_{0}$ (BJD$_{\rm{TDB}}$) & $2454508.97718 \pm 0.00022$\\
\hline                                   
\end{tabular}
\end{table}

The LD coefficients of the best-fitting model were compared to the theoretical values calculated from the tables from Claret \& Bloemen (\cite{Claret11}), and derived with the \textsc{EXOFAST} applet\footnote{http://astroutils.astronomy.ohio-state.edu/exofast/limbdark.shtml} (Eastman et al. \cite{Eastman12}). The coefficients were bilinearly interpolated from the tables assuming the host star effective temperature and surface gravity as given by Hebb et al.\ (\cite{Hebb}). The theoretical coefficients, $u_{\rm{t}}=0.288$ and $v_{\rm{t}}=0.317$ agrees with the derived values ($u=0.301^{+0.041}_{-0.040}$ and  $v=0.24^{+0.08}_{-0.08}$) within 1\,$\sigma$.

To check for any evolution of the system parameters, for instance as a result of a periastron rotation, the fitting procedure was repeated with $i_{\rm{b}}$, $a_{\rm{b}}/R_{*}$, and $R_{\rm{b}}/R_{*}$ allowed to vary between individual light curves. The linear and quadratic LD coefficients were allowed to vary around values derived in the previous run, under the Gaussian penalty defined by the derived errors. We found that none of these parameters exhibited a statistically significant variation. In addition, we compared individual mid-transit times derived in both approaches, and no discrepancy greater than 1\,$\sigma$ was found. 

The planet WASP-12~b is known to depart significantly from a spherical shape. The ratio of its polar and sub-stellar-point radii was found to be $\sim$1.14 (Budaj \cite{Budaj}). The observed planetary radius $R_{\rm{b}}$ is the cross section of the planet at inferior conjunction. It can be calculated from $a_{\rm{b}}/R_{*}$ and $R_{\rm{b}}/R_{*}$ assuming a value of the semi-major axis of $a_{\rm{b}}$ from Hebb et al. (\cite{Hebb}). The effective planetary radius, $R_{\rm{b}}^{\rm{eff}}$, which is more representative of the physical parameters of the planet, is greater. Following the numerical results of Budaj (\cite{Budaj}), we get $R_{\rm{b}}^{\rm{eff}} = 1.90 \pm 0.09$ $R_{\rm{Jup}}$. This value is identical to $1.90 \pm 0.09$ $R_{\rm{Jup}}$ determined by Maciejewski et al.\ (\cite{Maciejewski11}) and within 1-$\sigma$ of the results of Chan et al.\ (\cite{Chan}) and Southworth (\cite{Southworth12}). The mean planetary density was found to be $\rho_{\rm{b}}=0.206\pm0.043$ $\rho_{\rm{Jup}}$. This value is smaller than found by previous studies, because we have accounted for the effective planetary radius in our calculations, but is consistent with them within 1\,$\sigma$. The surface gravitational acceleration we find, $g_{\rm{b}}=9.8\pm0.4$ m~s$^{-2}$, is also noticeably smaller than previous studies, for the same reason.

A faint stellar companion to WASP-12 has been recently discovered at a separation of $\sim$$1\arcsec$ (Bergfors et al. \cite{Bergfors}). It is $\sim$$4$\,mag fainter in the $i\arcmin$ band and its spectral type is in the region of K0 to M0. As it is not resolved in any of our observations, its light contributes to the total flux of the system. This makes the transit depth shallower by 0.2--0.3 mmag, the precise amount depending on its spectral type. Due to this, we estimate the final planetary radius to be greater by 0.6--1.0\%, hence $\rho_{\rm{b}}$ and $g_{\rm{b}}$ will be slightly smaller than the values given above. 

The stellar parameters agree with previous studies. We find the stellar radius to be $R_{*}=1.63\pm0.07$ $R_{\odot}$ (assuming the value of $a_{\rm{b}}$ from Hebb et al. \cite{Hebb}), its mean density to be $\rho_{*}=0.315\pm0.007$ $\rho_{\odot}$, and its surface gravity to equal $\log g_{*} = 4.14\pm0.03$ (cgs units).

\subsection{Transit timing}\label{Timing}

\begin{figure*}
  \centering
  \includegraphics[width=18cm]{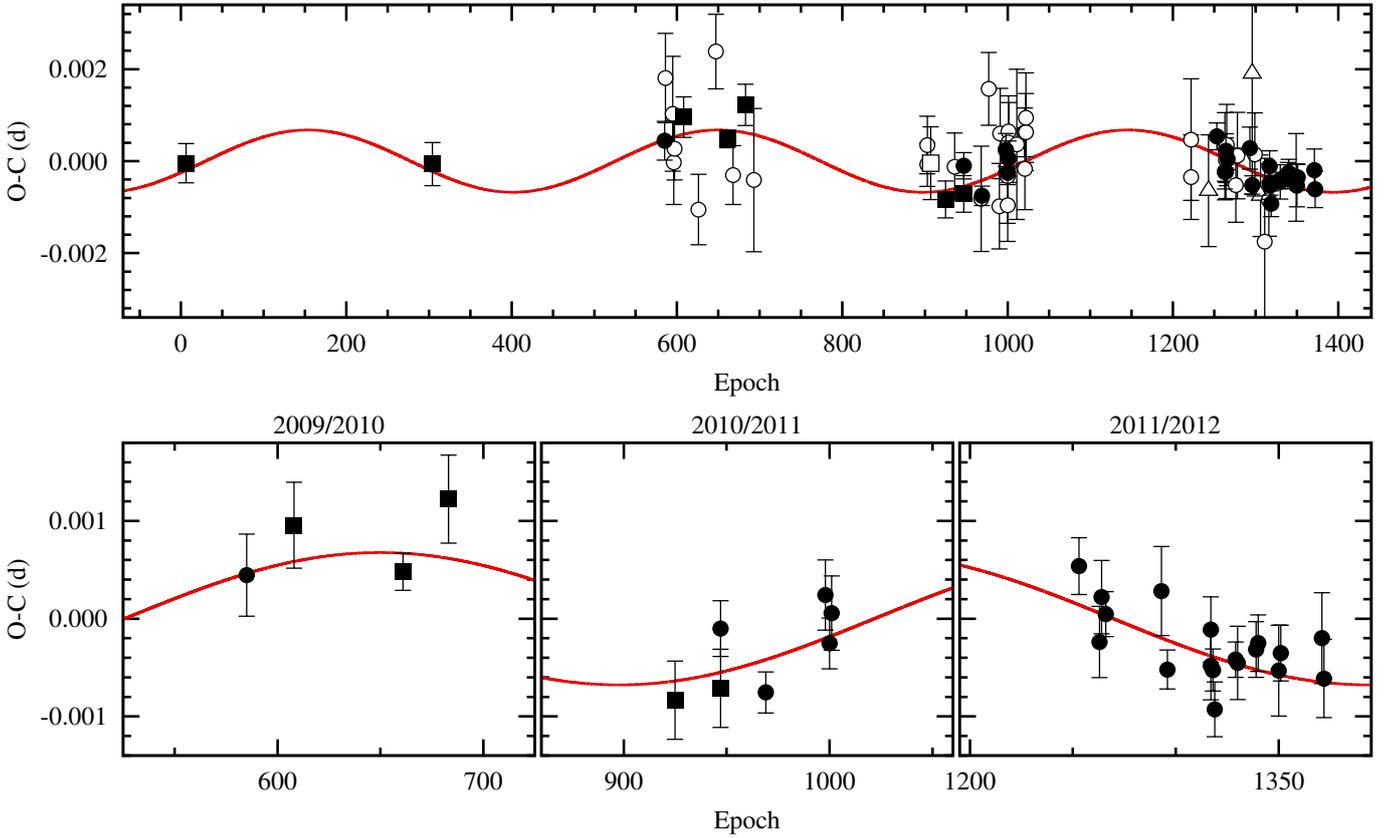}
  \caption{O--C diagram for WASP-12~b's transit timing. The filled circles and squares denote our observations and literature ones, respectively, with errors smaller than 40 s. The open symbols -- circles and squares for our and literature data, respectively -- mark points with timing errors greater than 40 s. The open triangles denote the amateur times included in this work. The postulated sinusoidal variation is sketched with a red line. Individual observing seasons, covered by our high-precision observations, are zoomed-in in the bottom. The literature data are taken from Hebb et al. (\cite{Hebb}), Chan et al. (\cite{Chan}), Maciejewski et al. (\cite{Maciejewski11}), Sada et al. (\cite{Sada}), and Cowan et al. (\cite{Cowan}).}
  \label{fig:oc}
\end{figure*}

The individual transit times were determined by fitting a template transit light curve with {\sc tap}. The fitting procedure started with initial parameters whose values are taken from the best-fitting model (Sect.~\ref{System}, Table \ref{table:pars}). They were allowed to vary in the MCMC analysis under Gaussian penalties whose scales were determined by the parameters' errors. This approach allowed best-model uncertainties to be included in a mid-transit time budget of each light curve. Ten chains of length of $10^5$ steps were used for each light curve. The first 10\% of iterations of each chain were discarded before calculating the parameter values and errors. The mid-transit times, converted to Barycentric Julian Dates in Barycentric Dynamical Time (BJD$_{\rm{TDB}}$, see Eastman et al.\ \cite{Eastman}), and their errors are given in Tables \ref{table:2} and \ref{table:3}. 

In addition, we re-analysed two datasets from Maciejewski et al.\ (\cite{Maciejewski11}) and obtained mid-transit times $2455230.40669\pm0.00019$ and $2455254.41871\pm0.00045$ BJD$_{\rm{TDB}}$. These values are well within error bars compared to the original mid-transit times of $2455230.40673\pm0.00011$ and $2455254.41887\pm0.00014$ BJD$_{\rm{TDB}}$, respectively. We also re-analysed a high-accuracy light curve acquired with the 2-m Liverpool Telescope and published in Hebb et al.\ (\cite{Hebb}). The mid-transit time for this early epoch (no.\ 6) was found to be $2454515.52496\pm0.00043$ BJD$_{\rm{TDB}}$ assuming times in original data are in HJD$_{\rm{UTC}}$\footnote{It is stated that the light curve is in JD$_{\rm{UTC}}$ (L.\ Hebb 2012, priv.\ comm.). However, the mid-transit time derived by us and the value published in Campo et al. (\cite{Campo11}) clearly indicate that the original light curve was converted into HJD$_{\rm{UTC}}$.}. The original mid-transit time was not published in Hebb et al.\ (\cite{Hebb}) but it is listed in Campo et al.\ (\cite{Campo11}) as $2454515.52542\pm0.00016$ BJD$_{\rm{TDB}}$. Both determinations differ by less than 1\,$\sigma$.

Mid-transit times of a single transit observed with different telescopes provide an empirical test of the timing accuracy and error estimates. We collected two or more light curves for twelve transits. Eight of them were observed with two telescopes and four with three telescopes. The median difference between individual times was found to be 0.4\,$\sigma$ and never exceeded 1\,$\sigma$. This finding supports the reliability and consistency of our timing survey. Furthermore, we compared mid-transit times derived with \textsc{tap} and \textsc{jktebop}. The latter code offers Monte Carlo (MC) simulations, a bootstrapping algorithm, and the residual-shift (prayer-bead) method to estimate the uncertainties of parameters. The largest of the errors from the methods were conservatively taken. Both codes give consistent results with the median difference between the determined times to midpoint equal to 13\,s or 0.13\,$\sigma$. The median of the ratios of the \textsc{tap} to the \textsc{jktebop} timing errors was found to be 1.33. This value includes partial light curves for which the \textsc{tap} code usually gives errors greater by a factor of up to four. 

A linear function was fitted to the transit times, to determine the orbital period $P_{\rm{b}}$ and the time of transit at cycle zero, $T_{0}$. Cycle zero was taken to be the epoch given by Hebb et al.\ (\cite{Hebb}). We fitted the transit times from new and literature data by the method of least-squares. The data point for epoch zero was skipped because it comes from a global fit based on a number of individual light curves (Hebb et al.\ \cite{Hebb}). We found $T_0 = 2454508.97718 \pm 0.00022$ BJD$_{\rm{TDB}}$ and $P_{\rm{b}} = 1.0914209 \pm 0.0000002$ d. The individual mid-transit errors were used as weights during the fitting. The reduced $\chi^2$ is $1.29$, which corresponds to a $p$-value of 0.06. Hence a null hypothesis assuming constant $P_{\rm{b}} $ could not be rejected. The O--C (Observed minus Calculated) diagram for transit timing is plotted in Fig.~\ref{fig:oc}.

\begin{figure}
  \centering
  \includegraphics[width=9cm]{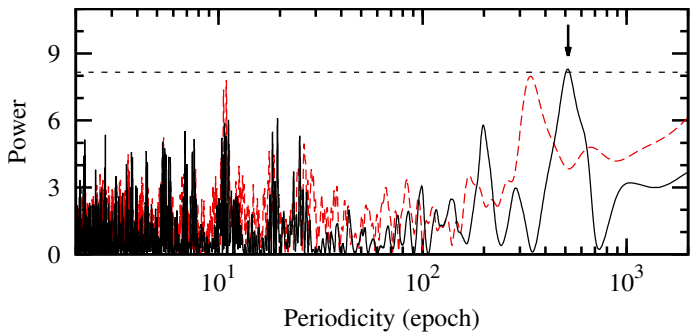}
  \caption{Periodogram of the O--C residuals (continuous black line) and spectral window (dashed red line, multiplied by 10 for clarity). The dashed vertical line shows the empirical FAP level of 1\%. The arrow marks the most significant peak which, as can be clearly seen, is not caused by the structure of gaps in the time domain. The shortest period investigated is given by the Nyquist frequency.}
  \label{fig:per}
\end{figure}

More interestingly, the same conclusion cannot be drawn if the mid-transit time sample is limited to the best-quality data. We now consider only those mid-transit times which have uncertainties below 40\,s (i.e. a half of the TTV amplitude suggested in Maciejewski et al.\ \cite{Maciejewski11}). The linear fit was repeated, resulting in $\chi^2=1.99$ and a $p$-value of $1.6\times10^{-3}$. This result allows the null hypothesis to be questioned. Moreover, a coherent pattern in the O--C diagram is noticeable (Fig.~\ref{fig:oc}). In the next step, a model consisting of a linear trend and a sinusoidal variation was fitted. After removing the linear component, a Lomb-Scargle periodogram (Lomb \cite{Lomb}, Scargle \cite{Scargle}) reveals a peak at a period of $\sim$510 epochs (Fig.~\ref{fig:per}). The false alarm probability (FAP) of this signal is 0.7\% and was determined empirically by a bootstrap resampling method. The procedure randomly permutes the O--C values retaining the original observing epochs and then calculates the Lomb--Scargle periodogram to determine the power of the strongest signal. After $10^5$ such trials, the FAP of the signal in the original dataset is taken to be the fraction of resampled periodograms that produced higher power than the original dataset. From the least-squares fit, we find that the possible TTV signal has a period of $500\pm20$ epochs and a semi-amplitude of $0.00068\pm0.00013$ d (Fig.~\ref{fig:oc}). The errors are taken from the covariance matrix of the fit. The reduced $\chi^2$ of this model is $0.91$.

\subsection{Two-planet model}\label{Model}

A TTV signal shown by the transiting planet may reflect gravitational perturbations driven by an undetected additional planet in the system (e.g.\ Ballard et al.\ \cite{Ballard11}). To check what planetary configurations may produce timing variations with the postulated periodicity, three-body simulations were performed with the \textsc{mercury} package (Chambers \cite{Chambers99}) employing the Bulirsch--Stoer integrator. A hypothetical additional planet of a probe mass of 40 $M_{\oplus}$ was put on orbits with semimajor axes ranging from 0.007 to 0.068 AU (orbital periods between 0.22 and 5.5 d, respectively) with a step equal to $10^{-6}$ AU. The semimajor axis of WASP-12~b and the mass of the central star were set to the values found by Hebb et al.\ (\cite{Hebb}). The eccentricity of the perturber, $e_{\rm{c}}$, was varied from 0.0 to 0.3 in steps of 0.05. Scenarios with circular and eccentric orbits ($e_{\rm{b}}$ equal to 0.02, 0.03 and 0.05) of the transiting planet were considered. The simulations assumed coplanar and edge-on orbits. The calculations covered 1500 orbital periods of the transiting planet, i.e.\ slightly greater than the time span of the timing observations. Lomb-Scargle periodograms were calculated for the synthetic O--C diagrams, to determine the dominant periodicity. Numerous solutions that reproduce the postulated TTV signal were identified. 

To alleviate the solution degeneracy, the publicly available RV measurements were reanalysed with the \textsc{Systemic Console} software (Meschiari et al.\ \cite{Meschiari09}). We used data from Hebb et al.\ (\cite{Hebb}), Husnoo et al.\ (\cite{Husnoo}), and Albrecht et al.\ (\cite{Albrecht}). The datapoints in transit phase ($\pm1.6$ h from predicted mid-transit times) were removed to avoid interference from the Rossiter--McLaughlin (RM) effect 
(Rossiter \cite{Rossiter}; McLaughlin \cite{McLaughlin}) 
that manifests itself as a distortion of the spectral lines of the host star during transit. Data published by Hebb et al.\ (\cite{Hebb}) were collected in the 2007/8 observing season with the SOPHIE spectrograph coupled with the 1.9-m telescope at the Observatoire de Haute Provence. We selected 17 RV measurements with errors between 9 and 13 m\,s$^{-1}$ for further analysis. Husnoo et al.\ (\cite{Husnoo}) presented further data acquired with SOPHIE in the 2008/9 and 2009/10 seasons. The spectra acquired on 2009 Mar 21 and 2009 Apr 13 have half the signal-to-noise ratio and noticeably greater errors than the other observations. This could indicate that these two exposures were affected by cloud, which, in turn, could introduce a systematic shift in the RV measurement (N.\ Husnoo 2012, priv.\ comm.). We noticed that one of these lower-quality points is an outlier at the 3\,$\sigma$ level regardless of the RV model under consideration. These two points were therefore not included in any further analysis. It is worth noting that keeping these outlying points does not quantitatively change the ensuing results but noticeably degrades the quality of the fits to the RVs. We also rejected measurements from 2009 Jan 17 when the signatures of the RM effect were observed. We finally used four and six datapoints from the 2008/9 and 2009/10 seasons, respectively. RVs reported by Albrecht et al.\ (\cite{Albrecht}) were acquired in the 2011/12 season with the HIRES spectrograph on the Keck telescope. We used 13 measurements in our study. 

The RV dataset from Husnoo et al.\ (\cite{Husnoo}) was split into two separate subsets according to observing season. We noticed that allowing a variation in the RV offsets between individual observing seasons results in significantly better fits. This effect may be caused by the low number of datapoints, the influence of additional bodies on long-term orbits in the system, or instrumental effects. The latter is a justifiable possibility because SOPHIE, when operated in the High Efficiency mode, has been  found to produce deviations up to several dozens of m\,s$^{-1}$ from a velocity zero point on a timescale of a few months (Husnoo et al.\ \cite{Husnoo}). 

In our analysis, the RV signature of tides raised in the host star by the transiting planet is not considered because its expected amplitude of 4.8 m\,s$^{-1}$ (Arras et al.\ \cite{Arras}) is smaller than the typical errors of the RV measurements.

\begin{figure}
  \centering
  \includegraphics[width=9cm]{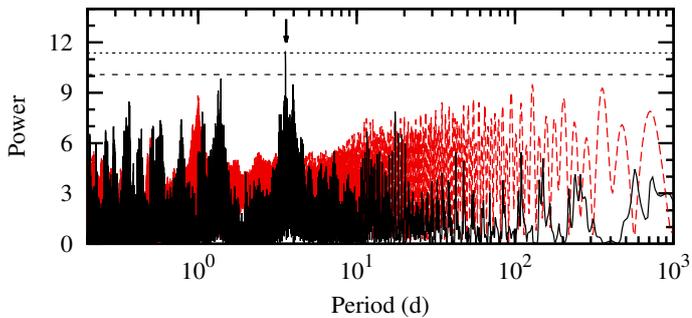}
  \caption{Periodogram of RV residuals for WASP-12~b on a Keplerian orbit (continuous black line) and the spectral window (dashed red line, multiplied by 10 for clarity). The arrow indicates the strongest peak at 3.58 d that is not associated with significant power in the spectral window. The shortest period is set by the radius of the star. The dashed and dotted horizontal lines show empirical FAP levels of 1\% and 0.1\%, respectively.}
  \label{fig:rvper}
\end{figure}

The periodogram of the RV residuals versus the best-fitting model for WASP-12~b on a circular Keplerian orbit (Fig.~\ref{fig:rvper}) shows a signal at approximately 3.6 d. Its empirical FAP was found to be 0.08\%, from $10^5$ trials with the bootstrap resampling method. A preliminary Newtonian RV modelling indicate that the residuals may be caused by an additional $\sim$$0.1$\,$M_{\rm{Jup}}$ planet on a non-circular orbit with an eccentricity $e_{\rm{c}}$ of between 0.2 and 0.3. Intriguingly, our three-body simulations performed with the \textsc{Mercury} code show that a system configuration for a $0.1$ $M_{\rm{Jup}}$ perturber may reproduce the postulated timing variations of WASP-12~b if $e_{\rm{c}}$ is close to 0.3 and $e_{\rm{b}}$ is between 0.03 and 0.05.    

\begin{figure}
  \centering
  \includegraphics[width=9cm]{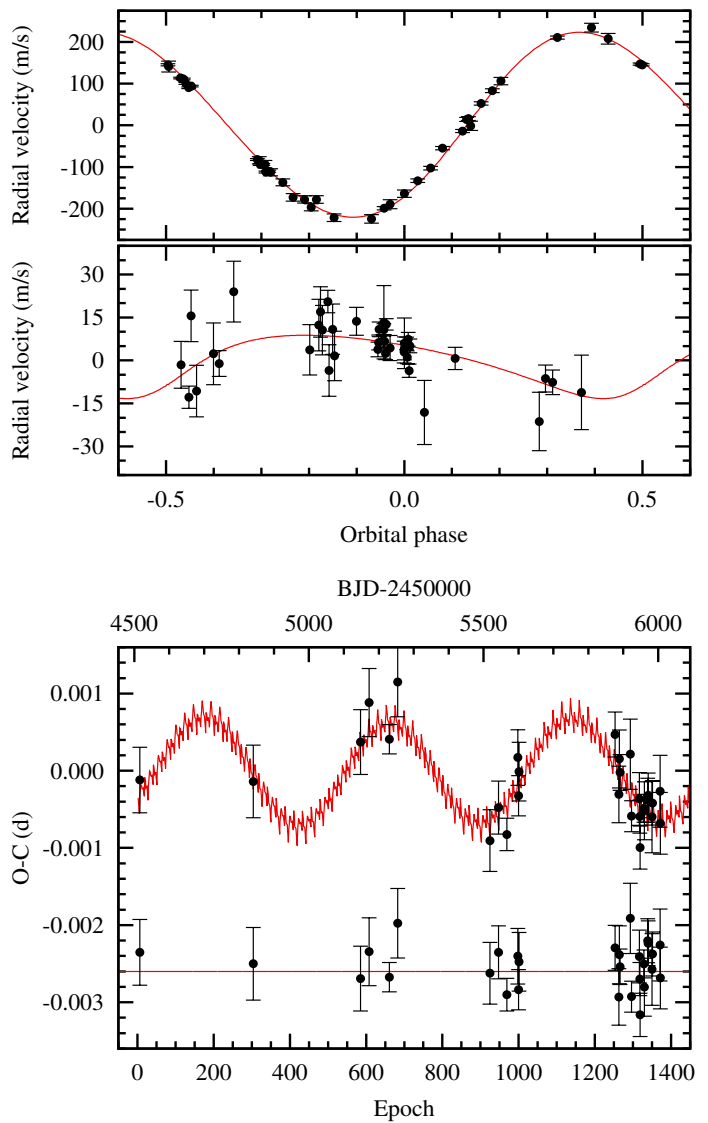}
  \caption{Joint RV and transit timing model for the WASP-12 system including two planets. {\em Upper}: the RV variation produced by WASP-12~b. {\em Middle}: the RV signal generated by the postulated planet WASP-12~c. {\em Bottom}: the O--C diagram for the transit timing of WASP-12~b with a modelled TTV signal and the residuals shifted by $-0.0026$ d for clarity.}
  \label{fig:model}
\end{figure}

To model the RV and transit timing datasets simultaneously, we used the \textsc{Systemic Console} with the Bulirsch--Stoer integrator and a precision parameter of $10^{-14}$. The times of RV observations were transformed to BJD$_{\rm{TDB}}$. To put additional constraints on the orbital parameters of WASP-12~b, the transit dataset was enhanced with five mid-occultation times from Campo et al.\ (\cite{Campo11}) and Croll et al.\ (\cite{Croll}). The light-travel time correction of 22.8\,s was taken into account. The L\'opez-Morales (\cite{Lopez}) mid-occultation point was not considered in the final iterations because of its much greater uncertainty and possible systematic timing offset (see Campo et al.\ \cite{Campo11} and Croll et al.\ \cite{Croll} for discussion). The orbital periods, masses, eccentricities, longitudes of periastron, and mean anomalies were allowed to vary for both planets. The offsets for individual RV subsets were also free parameters. The best-fitting model was found iteratively with the simulated annealing and Nelder-Mead minimisation algorithms. The joint $\chi^2$ statistic accounts for the contributions from the RV and transit time datasets, $\chi^{2}_{\rm{RV}}$ and $\chi^{2}_{\rm{tr}}$ respectively: 
\begin{equation}
  \chi^2= \frac{1}{N_{\rm{RV}}+N_{\rm{tr}}-N_{\rm{fit}}}[(N_{\rm{RV}}-N_{\rm{fit}})\chi^{2}_{\rm{RV}}+\lambda \chi^{2}_{\rm{tr}}]
\end{equation}
where $N_{\rm{RV}}$, $N_{\rm{tr}}$, and $N_{\rm{fit}}$ are the numbers of RV measurements, mid-transit times, and fitted parameters, respectively, and $\lambda$ is an arbitrary weight (Meschiari et al.\ \cite{Meschiari09}). The $\chi^{2}_{\rm{RV}}$ of the best-fitting model is approximately five times greater than $\chi^{2}_{\rm{tr}}$, which suggests that the RV errors are underestimated. To balance the significance of both data sets, $\lambda=5$ was specified in the final iterations. The MCMC method was used to determine parameter uncertainties. The MCMC chain was $10^6$ steps long, and the first $10^5$ configurations were discarded. The scale parameters were set empirically in a series of attempts to get the acceptance rate of the MCMC procedure close to the optimal value of $0.25$. For each parameter the standard deviation was taken as the final error estimate. The best-fitting parameters of the two-planet model are given in Table~\ref{table:2plpars}. Fig.~\ref{fig:model} shows the RV components from WASP-12~b and the postulated planet, as well as the modelled TTV signal.

\begin{table*}
\caption{Parameters of the two-planet WASP-12 system from the joint analysis of RVs and transit timing.} 
\label{table:2plpars}                       
\centering                  
\begin{tabular}{l c c}      
\hline\hline                
Parameter & Planet b & Planet c\\ 
\hline 
Orbital period, $P$ (d)                               & $1.0914281\pm0.0000015$ & $3.58484\pm0.00011$ \\
Minimum mass, $m \sin i$ ($M_{\rm{Jup}}$)             & $1.356\pm0.009$ & $0.098\pm0.009$ \\
Mean anomaly\tablefootmark{a}, $\lambda$ ($^{\circ}$) & $310.61\pm0.25$ & $184.6\pm1.1$ \\
Eccentricity, $e$                                     & $0.0447\pm0.0043$ & $0.284\pm0.002$ \\
Longitude of periastron, $\omega$ ($^{\circ}$)        & $274.44\pm0.03$ & $223.8\pm2.5$ \\
Semimajor axis, $a$ (AU)                              & $0.0229\pm0.0008$ & $0.0507\pm0.0018$ \\
Mass\tablefootmark{b}, $m$ ($M_{\rm{Jup}}$)           & $1.366\pm0.011$ & $0.098\pm0.009$ \\
\hline                                   
\end{tabular}
\\
\tablefoottext{a}{given for the initial epoch BJD$_{\rm{TDB}}$ 2454509.3871}\\
\tablefoottext{b}{assuming a coplanar system with the inclination derived from transit light curve modelling.}
\end{table*}

The RV datasets from the 2008/9 and 2009/10 seasons were found to be shifted against the 2007/8 dataset by $-16.4\pm3.3$ and $+1.1\pm2.9$ m~s$^{-1}$, respectively. The value of the joint $\chi^2$ was found to be 3.0 with individual components $\chi^{2}_{\rm{RV}}$ and $\chi^{2}_{\rm{tr}}$ equal to 2.5 and 0.4, respectively. For comparison, a single-planet model with a circular and eccentric orbit have joint $\chi^2$ values of 7.9 and 6.3, respectively. In the two-planet model, the final $rms$ of the RV dataset was reduced to 7.4 m~s$^{-1}$ from 12.5 m~s$^{-1}$ for a single-planet model with WASP-12~b on a circular orbit, and from 9.6 m~s$^{-1}$ for a single-planet model with WASP-12~b on an eccentric orbit.

\subsection{Dynamical stability}\label{Stability}

The relatively large eccentricity of the postulated planet WASP-12~c raises the
question of whether the system is dynamically stable. In this paper, we consider
only Newtonian, point-mass mutual interactions as a first-order model of the
dynamics of the system. The planet WASP-12~b has extremely short orbital
period, so it is not sufficient to study the long-term evolution of the
system with close-in companion WASP-12~c based only on the conservative
dynamics. Due to the close proximity of WASP-12~b to its host star, a full
dynamical model would need to account for general relativistic effects,
rotational and tidal deformations, and the tidal dissipation of energy (e.g.\
Mardling \& Lin \cite{Mardling2002}; Migaszewski \& Go\'zdziewski
\cite{Migaszewski2009}). If these effects are considered, the orbital
geometry and long-term evolution of the whole, mutually interacting system may be
strongly affected since its formation, possible migration, and tidal evolution.
We postpone this complex problem to another work.
 
A cursory inspection of the putative system parameters (Table
\ref{table:2plpars}) reveals that it is close to the 10b:3c mean-motion
resonance (MMR)\footnote{The $X$c:$Y$b notation means the ratio of orbital
periods of the inner and the outer planet, respectively.}. This justifies our
interest in the short-term dynamics of the system. We reconstruct the structure
of the phase-space in terms of the maximal Lyapunov exponent (Benettin et al.
\cite{Benettin1976}) which is expressed through the so called fast indicator
Mean Exponential Growth factor of Nearby Orbits (MEGNO, see e.g.\ Cincotta \&
Sim{\' o} \cite{Cincotta2000}; Cincotta et al.\ \cite{Cincotta2003};
{Go{\'z}dziewski et al.\ \cite{Gozdziewski2001} for details). This dynamical
characteristic helps us to classify given sets of initial conditions as
regular (leading to quasi-periodic evolution of the system, stable over
infinite period of time) or chaotic (leading to irregular phase-space
trajectories). The later might lead to geometric changes of the orbits rendering
the system unstable in the short-term timescale of $10^4$--$10^5$ outer periods
($P_{\rm{c}}$). By the design, the MEGNO indicator when converged to~2 over
a given integration time, guarantees orders of magnitude longer time-scale of
the dynamical (Lagrange) stability. However, one must be aware that we consider
a relatively narrow time window of the system evolution, restricted to the
conservative, Newtonian point-masses dynamics.

To visualise the orbital stability and dynamical neighbourhood of the WASP-12
system, we computed a high-resolution dynamical map in the
$(a_{\rm{c}},e_{\rm{c}})$--plane, see Fig.~\ref{fig:stability}, with the help of
our new CPU cluster software {\sc mechanic} (S{\l}onina et al.\
\cite{Slonina2012}). Each point (initial condition) was integrated for $10^5$
$P_{\rm{c}}$. Such a timescale is long enough to detect even weak, high-order
MMRs. Two MMRs (56c:17b and 33c:10b), which are the closest match to the
nominal position of the system (marked with the asterisk symbol), are labelled
on the scan. Thanks to the high resolution of the dynamical map, other,
even weaker MMRs are also identifiable. The 79c:24b MMR and the 89b:27c MMR
(both not labelled), which are between the 23b:7c MMR and 56c:17b~MMR, are
the closest to the nominal position of the system. Bearing in mind very
small error range of the semimajor axis of the outer planet in the dynamical
map, $~\sim 10^{-6}$~AU, (see an explanation below), we found that it might
encompass these resonances. Yet they are very narrow and weak due to their high
order. There is only a small probability that the system might be trapped in
such resonances. The error range of eccentricity safely separates the WASP-12
system from the zone of global chaos, visible roughly above $e_{\rm{c}} \sim
0.32$. This zone appears due to proximity to the orbits' crossing zone and
overlapping of the MMRs. 

Note that the dynamical map in Fig.~\ref{fig:stability} is computed for a
particular, nominal mass of the parent star and all orbital elements, besides
those ones constituting the map coordinates, are fixed at their best fit values.
In general, for a fixed nominal mass of the star, the semi-major axis derived as
osculating element at some initial epoch is essentially influenced by
uncertainties of the planetary masses and orbital periods in accord with the
IIId~Kepler law:
\begin{equation}
 a_{\rm{c}} = \left( \frac{k^2 (M_*+m_{\rm{c}}) P^2_{\rm{c}}}{(4\pi)^2} \right)^{(1/3)},
\end{equation}
as the most uncertain parameters (here $k$ is the Gaussian gravitational
constant), though the planetary and stellar mass relation is indirect due to the
$N$-body observational model. Therefore, the formal error of $a_{\rm{c}}$ in
the dynamical map should be considered as weakly dependent on the error of the
(fixed) stellar mass.

If the stellar mass is varied, the {\em relative} positions of the mean motion
resonances do not change, as they depend essentially on {\em the mass ratios}
\begin{equation}
 \frac{M_* + m_{\rm{c}}}{M_*+m_{\rm{b}}} \equiv 
 \left(1+\frac{m_{\rm{c}}}{M_*}\right)/\left(1+\frac{m_{\rm{b}}}{M_*}\right),
\end{equation}
and the ratios of the orbital periods. Hence, the dynamical map might be
considered to be representative for different masses of the parent star. We also
emphasize, following Michtchenko \& Malhotra
(\cite{MichtchenkoandMalhotra2004}), that in the limit of small planetary masses
and eccentricities, the Newtonian, conservative three-body dynamics depend on
the mass ratios and semi-major axes, and not on the individual masses and
semi-major axes as the system seems to be well separated from low-order MMRs
(see also Migaszewski \& Go\'zdziewski \cite{Migaszewski2009}).

\begin{figure}
\centering
\includegraphics[width=9cm]{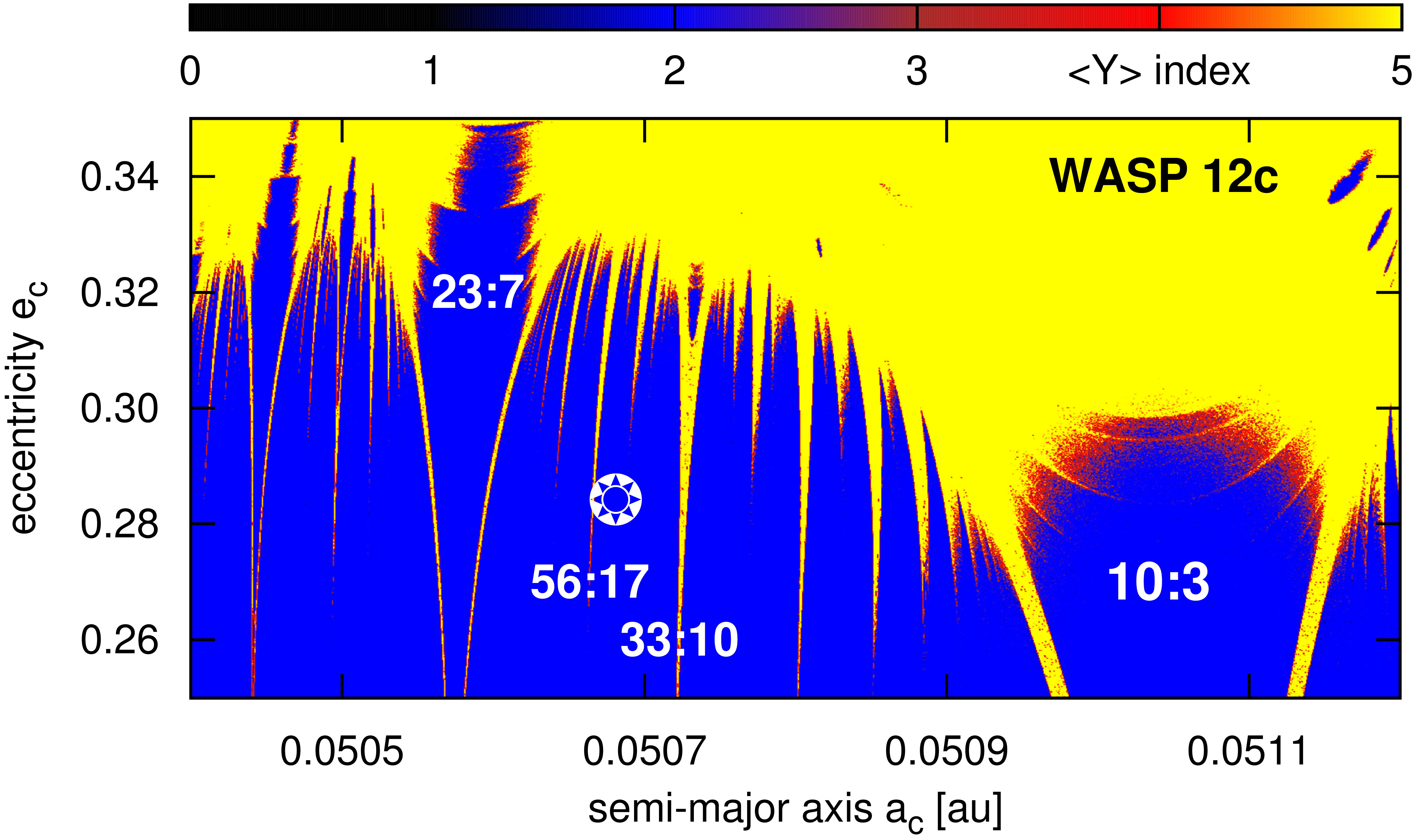}
\caption{Dynamical map of the WASP-12 system in the $(a_{\rm{c}},e_{\rm{c}})$--plane, computed in terms of the MEGNO fast indicator. These two orbital parameters are varied across the map, while other orbital elements and masses are kept at their nominal values from Table \ref{table:2plpars}. The star symbol, located at $a_{\rm{c}}=0.050681$ AU, indicates the nominal system. Colours encode the dynamical stability of the system: yellow is for strongly unstable configurations, and blue is for stable, quasi-periodic systems. A few relevant MMRs are labelled. The raw resolution of this map is $1440\times900$ datapoints. Each initial condition at the map has been integrated for $10^5$ outer orbits.}
\label{fig:stability}
\end{figure}


\section{Discussion}\label{Discuss}

The two-planet model predicts that the eccentricity of WASP-12~b might be non-zero. This, in turn, might affect the system parameters. The fitting procedure was repeated with non-zero $e_{\rm{b}}$ and $\omega_{\rm{b}}$ treated as fixed parameters, values of which were taken from Table~\ref{table:2plpars}. The noticeable difference was found for $i_{\rm{b}}$ and $a_{\rm{b}}/R_{*}$ which were found to be $82.51^{+0.53}_{-0.48}$ degrees and $2.93\pm0.02$, respectively. Compared to the circular-orbit scenario, the $i_{\rm{b}}$ and $a_{\rm{b}}/R_{*}$ determinations are smaller by 0.5 $\sigma$ and 2.5 $\sigma$, respectively. Adopting these values in the calculations results in $R_{\rm{b}}^{\rm{eff}}$ and $R_{*}$ greater by 4\% and 2\%, respectively, hence smaller values for $\rho_{\rm{b}}$, $g_{\rm{b}}$, $\rho_{*}$, and $\log g_{*}$. As the non-zero eccentricity of WASP-12~b is speculative, we settle for the scenario with the circular orbit in Sect.~\ref{System}.

Statistical studies of hot Jupiter candidates in the {\it Kepler} sample show that these planets are rare in multi-transiting systems (Latham et al.\ \cite{Latham}). The occurrence rate is below 5\%. The solitariness of hot Jupiters is postulated to be a result of the dynamical evolution of their planetary systems. On the other hand, there are systems with hot Jupiters accompanied by non-transiting planets on wide orbits, e.g.\ HAT-P-13 (Bakos et al. \cite{Bakos}), HAT-P-17 (Howard et al.\ \cite{Howard}), and Qatar-2 (Bryan et al.\ \cite{Bryan}). These additional planets have been discovered with the RV technique. Their gravitational influence on the transiting planets is too small to be detected through TTVs. Interestingly, single hot or warm Jupiter candidates which exhibit TTV signals have been found in the {\it Kepler} sample (Szab\'o et al.\ \cite{Szabo12}). The frequency of hot Jupiter candidates with periodic TTVs is estimated to be 18\%. More than half of these positive detections may be of the perturbative nature that allows scenarios with an additional close planet or exomoon to be considered. 

The ground-based timing of transiting hot Jupiters has brought no positive results so far, though preliminary detections have been announced. The characteristics of a spurious TTV signal for HAT-P-13~b are the most similar to the case of WASP-12~b.  P\'al et al.\ (\cite{Pal}) observed a 0.01\,d departure from a linear ephemeris. This finding was confirmed by Nascimbeni et al.\ (\cite{Nascimbeni}) who proposed a sinusoidal variation with a period of $\sim$1150\,d and semi-amplitude of 0.005\,d. Such a signal could not be caused by the known HAT-P-13~c planet, suggesting the existence of a third planetary component. Then, Fulton et al.\ (\cite{Fulton}) showed that all mid-transit times except the one from an earlier work by Szab\'o et al.\ (\cite{SzaboH13}) are consistent with a linear ephemeris. A similar conclusion was reached by Southworth et al.\ (\cite{SouthworthH13}) who found no evidence for the existence of periodic timing variations.

Bearing in mind the case of HAT-P-13~b, we observed various phases of the TTV signal postulated for WASP-12~b: a maximum (the 2009/10 season), two minima (the 2010/11 and 2011/12 seasons), and portions of ascent and descent trends. Furthermore, we selected mid-transit times of the highest quality for the final analysis. They come from different instruments and data reduction pipelines, thus systematic errors are expected to be minimised. The timing scatter between consecutive transits (up to four in the 2011/12 season) and runs (separated from each other by at least 10--12 epochs) was found to be much smaller than the amplitude of the postulated long-term variation.

Near-ultraviolet (near-UV) transmission spectroscopy during transits of WASP-12~b shows that the planet has an exosphere extending beyond the planet's Roche lobe (Fossati et al.\ \cite{Fossatib}). Moreover, these data show that transits begin earlier and are longer than at optical wavelengths. This phenomenon may be interpreted as a fingerprint of an accretion stream escaping from the planet through the inner Lagrangian point (Lai et al.\ \cite{Lai}). It is also explicable as the effect of a bow shock induced by the orbital motion of the planet with a magnetic field in the stellar coronal material (Vidotto et al.\ \cite{Vidotto}; Llama et al.\ \cite{Llama}). A characteristic of such dynamical interactions is temporal variation implied by the variable magnetic field of the central star (Vidotto et al.\ \cite{Vidotto11}). The changes in phases of the early ingress would affect the transit duration and hence the time of a mid-transit. This, in turn, would produce a TTV. The effect of the early ingress, however, is not expected to be detected in the optical bands. In Sect.~\ref{System} we analysed only complete transits, and found no variation in the transit duration. We also compared the values of the $rms$ in ingress, flat bottom, and egress phases of these light curves after phase folding. Changes in the transit duration would smear the ingress and egress data points in phase, causing a greater $rms$. We found the $rms$ values of 0.80, 0.76, and 0.77 mmag for ingress, flat bottom, and egress, respectively. These values are compatible with each other, so again we find no sign of transit duration variations (TDV). 

Other mechanisms which generate TTVs may also be rejected.  An effect induced by an exomoon is expected to produce a number of harmonic frequencies in the periodogram of the timing residuals (Kipping \cite{kipping09}). The interactions with a distant stellar or planetary component on a wide orbit would generate a long-period TTV through the light-travel-time effect (Montalto \cite{Montalto}). The amplitude of the TTV caused by the Applegate mechanism (Applegate \cite{Applegate}), which invokes quasi-periodic variations in the quadrupole moment of magnetically active stars, is expected to be a few seconds over the course of $\sim$10 years (Watson \& Marsh \cite{Watson}). The effects of transit parallax (Scharf \cite{Scharf}) and stellar proper motion (Rafikov \cite{Rafikov}) are to small to be detected for the WASP-12 system at present.

Croll et al.\ (\cite{Croll}) note that $e_{\rm{b}}$ is poorly constrained if $\omega_{\rm{b}}$ is close to $90^{\circ}$ or $270^{\circ}$, and this is the case for WASP-12~b. The literature determinations of this angle range from $270^{\circ}$ (Croll et al.\ \cite{Croll}) to $286^{\circ}$ (Hebb et al.\ \cite{Hebb}). Our two-planet model gives $\omega_{\rm{b}}=274\fdg44\pm0\fdg03$. We obtained values of $e_{\rm{b}}\cos \omega_{\rm{b}} = 0.0035 \pm 0.0002$ and $e_{\rm{b}}\sin \omega_{\rm{b}} = -0.0446\pm0.0034$, and both of them are clearly far from zero. For the latter parameter, the consistency with the literature determinations is between 1.1 and 1.3 $\sigma$ for Campo et al.\ (\cite{Campo11}) and Croll et al.\ (\cite{Croll}), respectively. In the case of $e_{\rm{b}}\cos \omega_{\rm{b}}$, the values are seemingly discrepant if they are compared to our value obtained for the initial epoch 2454509 BJD. 

Our model neglects non-Keplerian sources of the apsidal precession including general relativistic effects, the quadrupole field caused by rotational flattening of the star and planet, and tidal potentials of the star and planet. This approach is justified because these effects are usually small, and hence generate precession rates much lower than the precession due to the second planet. However, it has been shown that the precession rate due to the tidal potential of WASP-12~b may be non-negligible (Ragozzine \& Wolf \cite{Ragozzine}). In this case, the precession period is predicted to be several years or decades, depending on the response of the planetary body to a tidal potential.

WASP-12 is a quiet star with no signatures of a structured magnetic field (Fossati et al.\ \cite{Fossatia}) and a chromospheric activity index $\log R^{\prime}_{\rm{HK}}=-5.5$ (Knutson et al.\ \cite{Knutson}; Krej\v{c}ov\'a \& Budaj \cite{Krejcova}). To estimate the expected amount of stellar RV jitter generated by the non-uniform convection (Saar et al.\ \cite{Saar}), we analysed the results of Isaacson \& Fischer (\cite{Isaacson}) who studied RV stability for more than 2600 main sequence and subgiant stars with spectral types between F5 and M4. We selected a sample of 310 stars whose colour index $B-V$ and activity index $\log R^{\prime}_{\rm{HK}}$ are close to these of WASP-12. As WASP-12 has $B-V=0.6$ mag we considered stars with $0.5 \leq B-V \leq 0.7$ mag, and with $\log R^{\prime}_{\rm{HK}} \leq -5.0$. The median jitter was found to be 2.6 m\,s$^{-1}$. The stellar jitter of WASP-12 is too high in a single-planet model, being 12.5 and 6.7 m\,s$^{-1}$ in the cases of circular and eccentric orbits for WASP-12~b, respectively. In the two-planet model, the jitter was found to be 2.7 m\,s$^{-1}$, which is consistent with the expected value from the Isaacson \& Fischer (\cite{Isaacson}) survey. To verify the jitter of WASP-12, we used the Albrecht et al.\ (\cite{Albrecht}) RV dataset which contains precise measurements from a single night. In such a short time interval of a few hours, the RV variation is generated by stellar jitter, planets, and possible instrumental trends. The latter two components can be temporarily approximated with a single Keplerian orbit. This approach yields a jitter of 2.7 m\,s$^{-1}$ for WASP-12, in agreement with the expected value.

The radii of known transiting planets with masses close to 0.1 $M_{\rm{Jup}}$ range from 0.21 $R_{\rm{Jup}}$ (Kepler-22~b, Borucki et al.\ \cite{Kepler22}) to 0.73 $R_{\rm{Jup}}$ (Kepler-35(AB)~b, Welsh et al. \cite{Kepler35}). If the postulated planet transits the host star, the corresponding flux drop is between 0.02\% and 0.2\%. If the planet is bloated to 1~$R_{\rm{Jup}}$, the expected transit depth is 0.4\%, so such a signal could be detected with a 1-m class telescope. The geometric transit probability for an eccentric orbit (Kane \& von Braun \cite{Kane}) of the postulated planet\footnote{Here we assume a planetary radius of 0.7 $R_{\rm{Jup}}$.} was found to be $13\pm4$\%, i.e.\ almost three times lower than for WASP-12~b. If the system is coplanar, the additional planet could produce grazing transits. 

The co-planarity of the system was found to be supported by numerical experiment in which the mutual inclination of both planets, $\Delta i$, was varied from 0 to $30\degr$ with a step size of $2\degr$ for $\Delta i<10\degr$ and $5\degr$ for $10\degr \leq \Delta i \leq 30\degr$. The goodness of the fit for the two-planet model was found to decrease with increasing $\Delta i$. Scenarios with $\Delta i \leq 10\degr$ result in $\chi^{2}_{\rm{tr}}<1$, so the two-planet WASP-12 system is likely to be coplanar. This finding testifies in favour of quiescent mechanisms by which planets are transported inwards with the planets' orbital orientation preserved (i.e.\ interactions between planets and protoplanetary discs). On the other hand, the obliquity of the host star (the sky-projected angle between the stellar spin and a planetary orbital axis) was found to be $59^{+15}_{-20}$ degrees (Albrecht et al.\ \cite{Albrecht}). Such a high value suggests that WASP-12~b migrated inwards by a process changing the relative orientation of the stellar and planetary axes (i.e.\ planet--planet interactions). It cannot be completely ruled out that the axial misalignment has a primordial origin unrelated to the evolution of the planetary system. It has been shown that an interaction between a magnetosphere of a young magnetic star and circumstellar disc may produce a significant spin--orbit misalignment (Lai et al.\ \cite{Lai11}).


\section{Conclusions}\label{Conclude}

Transit timings of WASP-12~b and precise velocities of its host star give marginally significant support for the presence of the second planet, when both datasets are considered separately. However, a joint analysis results in a two-planet model which better explains observations than single-planet scenarios. This approach removes the degeneracy of solutions which is typical of the TTV method. Our finding allows us to advance the hypothesis that WASP-12~b is not the only planet in the WASP-12 system, and that there is an additional 0.1 $M_{\rm{Jup}}$ body on a 3.6-d period eccentric orbit. The dynamical analysis of the two-planet system shows that it is located in a relatively wide stable zone. This reinforces our hypothesis. However, we note that the putative planetary companion remains under the threshold of reliable detection. Thus, further photometric and spectroscopic follow-up observations of high precision are needed to shed new light on the architecture of the WASP-12 system.


\begin{acknowledgements}

We are grateful to Dr.\ Simon Albrecht for making radial velocity data available to us at the stage of manuscript publication and to Dr.\ Leslie Hebb for sharing the Liverpool Telescope light curve. We  thank C.\ Adam and N.\ Pawellek for trying to obtain more data. We also thank the anonymous referee for suggestions which improved the manuscript.

GM, {\L}B, GN and AN acknowledge the financial support from the Polish Ministry of Science and Higher Education through the Iuventus Plus grants IP2010 023070 and IP2011 031971. GM and AN acknowledge funding from the European Community's Seventh Framework Programme (FP7/2007-2013) under grant agreement number RG226604 (OPTICON). DD acknowledges the financial support of the projects DO 02-362, and DDVU 02/40-2010 of the Bulgarian National Science Fund. DD and VP gratefully acknowledge observing grant support from the Institute of Astronomy and NAO, Bulgarian Academy of Sciences. 
SR, MK, RE, AB, CG, TR, and TOBS together with RN
would like to acknowledge support from the German
national science foundation Deutsche Forschungsgemeinschaft
(DFG) in grants NE 515 / 33-1, 33-2, 34-1, 32-1, 30-1,
and 36-1; projects NE 515 / 33-1, 33-2, and 34-1 are
part of the DFG Priority Programme SPP 1385 on
{\em The first 10 Myr of the Solar System}.
{\L}B acknowledges travel funds from the PAN/BAN exchange and joint research project ``Spectral and photometric studies of variable stars''.
RE also thanks the Abbe-School of Photonics for support.
CM thanks DFG for support through SCHR 665 / 7-1.
KG is supported by the Polish Ministry of Science and Higher Education No.\ N/N203/402739 and ``HPC Infrastructure for Grand Challenges of Science and Engineering'' (POWIEW) co-financed by the European Regional Development Fund under the Innovative Economy Operational Programme. TCH gratefully acknowledges financial support from the Korea Research Council for Fundamental Science and Technology (KRCF) through the Young Research Scientist Fellowship Program. TCH, JWL and CUL acknowledge financial support from KASI grant number 2012-1-410-02. They also expresses their thanks to the BOAO/LOAO support astronomers Hyung-Il, Sang-Min and Jae-Hyuck Youn. 
CG and MM acknowledge support from DFG in MU 2695/13-1.
MMH, NT, and RN would like to thank DFG for support in the collaborative research project (Sonderforschungsbereich SFB TR 7) on gravitational wave astronomy, sub-projects C2 and B9. NT would like to thank the Carl-Zeiss-Foundation for a scholarship. 
JS acknowledges financial support from STFC in the form of an Advanced Fellowship. KT is financially supported by the Grants-in-Aid for the Scientific Research by the Ministry of Education, Culture, Sports, Science and Technology in Japan (No.~23540277).  
KS and GyMSz acknowledge support through the Lend\"ulet-2009 Young
Researchers' Programme of the Hungarian Academy of Sciences, the
HUMAN MB08C 81013 grant of the MAG Zrt and the Hungarian OTKA Grants K-83790 and K-104607. 
GyMSz is also supported by J\'anos Bolyai Research Scholarship of the Hungarian Academy of Sciences. The contribution of MV was supported by the Slovak Research and Development Agency under the contract No.\ APVV-0158-11. MV also thanks for the support to the project VEGA 2/0094/11. 

We would like to acknowledge financial support from the Thuringian
government (B 515-07010) for the STK CCD camera used in this project.

The data presented here were obtained in part with ALFOSC, which is provided by the Instituto de Astrof\'{\i}sica de Andaluc\'{\i}a (IAA) under a joint agreement with the University of Copenhagen and NOTSA.

\end{acknowledgements}


\Online

\end{document}